\documentclass[journal]{IEEEtran}
\usepackage{latexsym,amsfonts,amssymb,amsthm, setspace, verbatim,cite,mathrsfs,graphicx,bm,stfloats, hyperref, tikz,xcolor,soul}
\usepackage[cmex10]{amsmath}
\newcommand{\be}{\begin{equation}}
\newcommand{\ee}{\end{equation}}
\newcommand{\ben}{\begin{equation*}}
\newcommand{\een}{\end{equation*}}
\newcommand{\mc}{\mathcal}
\newtheorem{lem}{Lemma}

\newtheorem{defi}{Definition}
\newtheorem{thm}{Theorem}

\newtheorem{claim}{Claim}

\newtheorem{corr}{Corollary}

\newcommand{\e}{\epsilon}
\newcommand{\tmb}[1]{\tilde{\mathbf{#1}}}

\begin{document}
\title{An Achievable Rate Region for the Broadcast Channel with Feedback}
\author{Ramji Venkataramanan,~\IEEEmembership{Member,~IEEE,}
S.~Sandeep Pradhan,~\IEEEmembership{Member,~IEEE}

\thanks{ This work was supported by NSF grants  CCF-0915619 and CCF-1111061. It was presented in part at the 2010 IEEE International Symposium on Information Theory, held in   Austin, TX.}%
\thanks{R.~Venkataramanan is with the Department of Engineering, University of Cambridge, Cambridge CB2 1PZ, UK (e-mail: ramji.v@eng.cam.ac.uk).}%
\thanks{S. Sandeep Pradhan is with the Department of Electrical Engineering and
Computer Science, University of Michigan, Ann Arbor, MI 48109, USA (e-mail: pradhanv@eecs.umich.edu).}
}
\maketitle

\begin{abstract}
A single-letter achievable rate region is proposed for the two-receiver discrete memoryless broadcast channel with generalized feedback. The coding strategy
involves block-Markov superposition coding using  Marton's coding scheme for the broadcast channel without feedback as the starting point. If the message rates in the Marton scheme are too high  to be decoded at the end of a block, each receiver is left with a list of messages compatible with its output. Resolution information is sent in the following block to enable each receiver to resolve its list. The key observation is that the resolution information of the first receiver is correlated with that of the second. This correlated information is efficiently transmitted via joint source-channel coding, using ideas similar to the Han-Costa coding scheme. Using the result, we obtain an achievable rate region for the stochastically degraded AWGN broadcast channel with noisy feedback from only one receiver. It is shown that this region is strictly larger than the no-feedback capacity region.
\end{abstract}


\section{Introduction}
\label{sec:intro}
\begin{figure}
\includegraphics[height=1.57in]{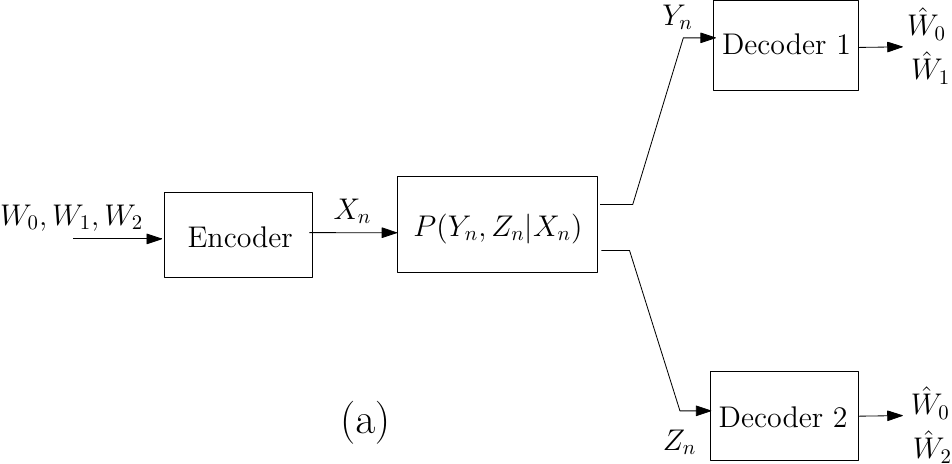} \\ \\ \\
\includegraphics[height=1.5in]{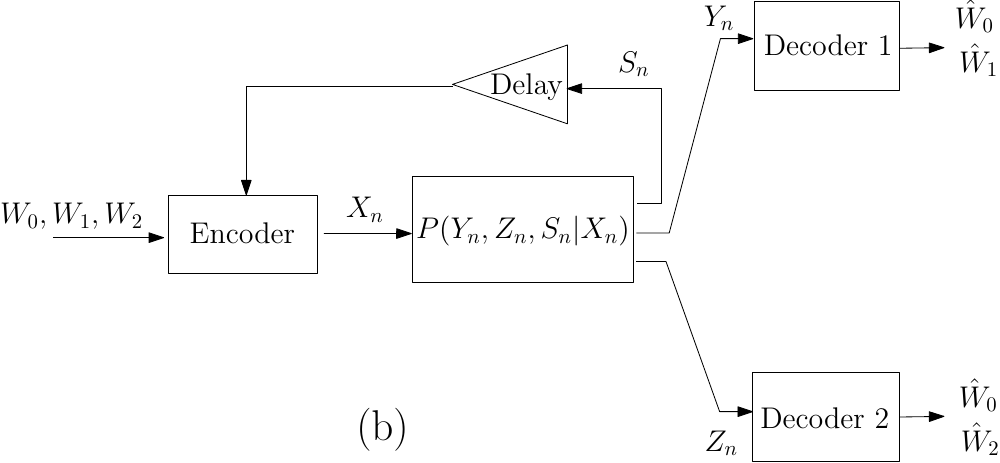}
\caption{The discrete memoryless broadcast channel with a) no feedback b) generalized feedback.}
\vspace{-1pt}
\label{fig:bc_fb}
\end{figure}

\IEEEPARstart{T}{he} two-receiver discrete memoryless broadcast channel (BC) is shown in  Figure \ref{fig:bc_fb}(a).
The channel has one transmitter which generates a channel input  $X$, and two receivers which receive $Y$ and $Z$, respectively.
The channel is characterized by a conditional law $P_{YZ|X}$. The transmitter wishes to communicate information   simultaneously to the receivers
at rates $(R_0,R_1, R_2)$, where $R_0$ is the rate of the common message, and $R_1, R_2$ are the rates of the private messages of the two receivers.
This channel has been studied extensively. The largest known set of achievable rates for this channel {without}
feedback is due to  Marton \cite{Marton79}. Marton's rate region is equal to the capacity region in all cases where
it is known.  (See \cite{Cover98}, for example, for a list of such channels.)

Figure \ref{fig:bc_fb}(b) shows a BC with generalized feedback. $S_n$ represents the feedback signal available at the transmitter at time $n$. This model includes noiseless feedback from both receivers $(S_n = (Y_{n}, Z_{n}))$, partial feedback $(S_n = Y_{n})$ as well as noisy feedback $(S_n = Y_{n} + \text{ noise})$.  El Gamal showed in \cite{ElG78} that feedback does not enlarge the
capacity region of a physically degraded BC. Later, through a simple example, Dueck \cite{Dueck80}  demonstrated
that feedback can strictly improve the capacity region of a general BC. For the stochastically degraded AWGN broadcast
channel with noiseless feedback, an achievable rate region larger
than the no-feedback capacity region was established in
\cite{OzarowLeung84}, and more recently, in \cite{Bhaskaran08}. A finite-letter achievable rate region (in terms of directed information) for the discrete memoryless BC with feedback was obtained by Kramer \cite{Kramer03}; using this characterization,  it was shown that  rates strictly outside the no-feedback capacity region could be achieved for the binary symmetric BC with noiseless feedback.

In this paper,  we establish a single-letter achievable rate region for the memoryless BC with generalized feedback. We use the proposed region to compute achievable rates for the stochastically degraded AWGN BC with noisy feedback from one receiver, and show that rates strictly outside the no-feedback capacity region can be achieved.

\begin{figure}
\centering
\includegraphics[width=2.9in]{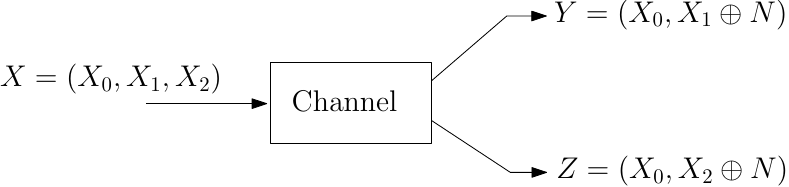}
\caption{The channel input is a binary triple $(X_0,X_1,X_2)$.  $N \sim$ Bernoulli$(\frac{1}{2})$ is an independent noise variable.}
\vspace{-1pt}
\label{fig:dueckbc}
\end{figure}

Before describing our coding strategy, let us revisit the example from \cite{Dueck80}.  Consider the BC in Figure \ref{fig:dueckbc}. The channel input is a binary triple $(X_0,X_1,X_2)$. $X_0$ is transmitted cleanly to both receivers. In addition, receiver $1$ receives $X_1 \oplus N$ and receiver $2$ receives $X_2 \oplus N$, where $N$ is an independent binary Bernoulli$(\frac{1}{2})$ noise variable. Here, the operation $\oplus$ denotes the modulo-two sum. Without feedback, the maximum sum rate for this channel is $1$ bit/channel use, achieved by using the clean input $X_0$ alone. In other words, no information can be reliably
transmitted through inputs $X_1$ and $X_2$.

Dueck described a simple scheme to achieve a greater sum rate using feedback. In the first channel use, transmit one bit to each receiver $i$ through $X_i, \: i=1,2$. Receivers $1$ and $2$ then receive $Y=X_{1} \oplus N$ and $Z=X_2 \oplus N$, respectively, and cannot recover $X_i$. The transmitter  learns $Y,Z$ through feedback and can compute $N =Y\oplus X_1 =Z \oplus X_2$.  For the next channel use, the transmitter sets $X_0=N$. Since $X_0$ is received noiselessly by both receivers, receiver $1$ can now recover $X_1$ as  $Y \oplus N$. Similarly, receiver $2$ reconstructs $X_2$ as  $Z \oplus N$. We can repeat this idea over several transmissions: in each channel use, transmit a fresh pair of bits  (through $X_1,X_2$) as well as  the noise realization of the previous channel use (through $X_0$). This yields a sum rate of $2$ bits/channel use. This is, in fact, the sum-capacity of the channel since it equals the cut-set bound $\max_{P_X} I(X;YZ)$.

The  example suggests a natural way to exploit feedback in a broadcast channel.  If we transmit a block of information at rates outside the no-feedback capacity region, the receivers cannot uniquely decode their messages at the end of the block. Each receiver now has a list of codewords that are jointly typical with its channel output. In the next block, we attempt to resolve these lists at the two receivers. The key observation is that the resolution information needed by receiver $1$ is in general \emph{correlated} with the resolution information needed by receiver $2$. The above example is an extreme case of this: the resolution information of the two receivers is identical, i.e., the correlation is perfect!

In general,  the two receivers'  resolution information are not perfectly correlated, but can still be transmitted over the BC more efficiently  than  independent information. This is analogous to transmitting correlated sources over a BC using joint source-channel coding \cite{HanCosta87,KramerNair09,MineroKim09,ChoiP08,KangKramer08}. At the heart of the proposed coding scheme is a way to represent the resolution information  of the two receivers as a pair of correlated sources, which is then transmitted efficiently in the next block using joint source-channel coding, along the lines of \cite{HanCosta87}. We repeat this idea over several blocks of transmission, with each
block containing independent fresh information superimposed over correlated resolution information for the previous block.

The following are the main contributions of this paper:
\begin{itemize}
\item We obtain a single-letter achievable rate region for the discrete memoryless BC with generalized feedback. The proposed region contains three extra random variables in addition to those in Marton's rate region.

\item Using a simpler form of the rate region with only one extra random variable, we compute achievable rates for the AWGN broadcast channel with noisy feedback. It is shown that rates outside the no-feedback capacity region can be achieved even with noisy feedback from only one receiver. This is the first characterization of achievable rates for the AWGN BC with noisy feedback at finite SNR, and is in contrast to the finding in \cite{PrelogDouble} that noisy feedback does not increase the prelog of the sum-capacity as the SNR grows asymptotically large.

One feature of the proposed region is that it includes the case where there a common message to be transmitted to both receivers, in addition to their private messages. The previously known schemes for the AWGN BC with noiseless feedback \cite{OzarowLeung84,Bhaskaran08} assume that there is no common message.
\end{itemize}

 At the conference where our result was first presented \cite{RP10}, another rate region for the BC with feedback was proposed independently by Shayevitz and Wigger \cite{WiggerShay}. Though a direct comparison of the two regions does not appear feasible, we show that the rates for the examples presented in \cite{WiggerShay, WiggerShayJourn} can also be obtained using the proposed region.

\emph{Notation}: We use uppercase letters to denote random variables, lower-case for their realizations and calligraphic notation for their alphabets. Bold-face notation is used for random vectors. Unless otherwise stated, all vectors have length $n$. Thus $\mathbf{A} \triangleq A^n \triangleq (A_1,\ldots,A_n)$  represents a random vector, and $\mathbf{a} \triangleq a^n  \triangleq (a_1,\ldots,a_n)$ a realization. The $\epsilon$-strongly typical set of block-length $n$
of a random variable with distribution   $P$ is denoted $\mc{A}_{\epsilon}^{(n)}(P)$. $\delta(\e)$ is used to denote a generic positive function of $\e$
that goes to zero as $\e \to 0$. Logarithms are with base $2$, and entropy and mutual information are measured in bits. For $\alpha \in (0,1)$, $\bar{\alpha} \triangleq 1-\alpha$. $\oplus$ denotes modulo-two addition.

In the following, we give an intuitive description of a two-phase coding scheme for communicating over a BC with noiseless feedback. We will use the  notation $\sim$ to indicate the random variables used in the first phase.  Thus $(\tilde{Y},\tilde{Z})$ denotes the channel output pair for the first phase, and $(Y,Z)$  the output pair for the second phase.  We start with Marton's coding strategy for the discrete memoryless BC without feedback. The  message rates of the two receivers are assumed to lie outside Marton's achievable rate region.
Let $\tilde{U},\tilde{V}$, and $\tilde{W}$ denote the auxiliary random variables used to encode the information. $\tilde{W}$ carries the information meant to be decoded at both receivers. $\tilde{U}$ and $\tilde{V}$ carry the rest of the information meant for the receivers $1$ and $2$, respectively.  The $\tilde{U}$- and $\tilde{V}$-codebooks are  constructed by randomly sampling the $\tilde{U}$- and $\tilde{V}$-typical sets, respectively. Let $\tilde{\mathbf{U}}$, $\tilde{\mathbf{V}}$ and $\tilde{\mathbf{W}}$ denote the three random codewords chosen by the transmitter.  The channel input vector $\tilde{\mathbf{X}}$ is obtained by `fusing' the triple $(\tilde{\mathbf{U}},\tilde{\mathbf{V}},\tilde{\mathbf{W}})$.

Since the rates lie outside Marton's region, the receivers may not be able to decode the information contained in $\tilde{U}, \tilde{V}$, and $\tilde{W}$. Instead, they can only produce a list of highly likely codewords  given their  respective channel output vectors. At the first decoder, this list is formed by collecting all $(\tilde{U},\tilde{W})$-codeword pairs that are jointly typical with the channel output. A similar list of $(\tilde{V},\tilde{W})$-codeword pairs is formed at the second receiver.  Note that even with feedback, the total transmission rate of the BC cannot exceed the capacity of the point-to-point channel with input $\tilde{X}$ and outputs $(\tilde{Y},\tilde{Z})$ (since the channel is memoryless). Hence, given \emph{both} channel output vectors $(\tilde{\mathbf{Y}},\tilde{\mathbf{Z}})$, the posterior probability of the codewords will be concentrated on the transmitted codeword triple.

At the end of the first phase, the feedback vector $\tilde{\mathbf{S}}$ is available at the encoder. In the second phase, we treat $(\tilde{\mathbf{U}},\tilde{\mathbf{W}})$ as the source of information to be transmitted to the  first  decoder, and $(\tilde{\mathbf{V}},\tilde{\mathbf{W}})$ as the source of information to be transmitted to the second  decoder.  The objective in the second phase is to communicate these two correlated pairs over the BC, while treating $\tilde{\mathbf{S}}$ as source state information and $\tilde{\mathbf{Y}}$ and $\tilde{\mathbf{Z}}$ as side-information available at the two receivers.  This is accomplished using a joint source-channel coding strategy. Transmission of correlated information over a BC has been addressed in \cite{HanCosta87,ChoiP08}.

In the Han-Costa framework \cite{HanCosta87}, the correlated  information is modeled as a pair of memoryless sources characterized by a fixed single-letter distribution. The pair of sources is first covered using codebooks constructed from auxiliary random variables; the covering codewords are then transmitted over the BC using Marton coding. The current setup differs from \cite{HanCosta87} in two ways. First, the correlated information  given by $(\tilde{\mathbf{U}},\tilde{\mathbf{W}})$ and $(\tilde{\mathbf{V}},\tilde{\mathbf{W}})$ does not exhibit a memoryless-source-like  behavior. This is because the vectors $\tilde{\mathbf{U}}$, $\tilde{\mathbf{V}}$ and $\tilde{\mathbf{W}}$ come from
codebooks.  However, when the codewords are sufficiently long and are chosen randomly, $(\tilde{\mathbf{U}}, \tilde{\mathbf{V}},\tilde{\mathbf{W}})$ will
be jointly typical  and can be covered using auxiliary codebooks similar to \cite{HanCosta87}.  The second difference from \cite{HanCosta87} is the presence of source state information $\tilde{\mathbf{S}}$ and side-information $\tilde{\mathbf{Y}}$ and $\tilde{\mathbf{Z}}$ available at receivers $1$ and $2$, respectively. We handle this by extending both the covering and channel coding steps of the Han-Costa scheme to incorporate the side-information.
Thus at the end of the second phase, the decoders are able to decode their respective messages.

We will superimpose the two phases using a block-Markov strategy.  The overall transmission scheme has several blocks, with fresh information entering in each block being decoded in the subsequent block. The fresh information gets encoded in the first phase, and is superimposed on the second phase which corresponds to  information that entered in the previous block.

It turns out that the performance of such a scheme cannot be directly captured by single-letter information quantities. This is because the state information,
given by the channel outputs of all the previous blocks, keeps accumulating, leading to a different joint distribution of the random variables
in each block. We address this issue by constraining the distributions used in the second phase (Definition \ref{def:joint_dist}) so that in every block, all the sequences follow a stationary joint distribution. This results in a first-order stationary Markov process of the sequences across blocks.

The rest of the paper is organized as follows. In Section \ref{sec:problem}, we define the problem formally and state the main result, an achievable rate region for BC with generalized feedback. We outline  the proof of the coding theorem in Section \ref{sec:coding_scheme}.
In Section \ref{sec:example}, we use the proposed region to compute achievable rates for the AWGN BC with noisy feedback. We also compare our region with the one proposed by Shayevitz and Wigger.  The formal proof of the coding theorem  is given in Section \ref{sec:proof}. Section \ref{sec:conc} concludes the paper.
\section{Problem Statement and Main Result}\label{sec:problem}
A two-user discrete memoryless broadcast channel with generalized feedback is a quintuple $(\mathcal{X},\mathcal{Y},\mathcal{Z},\mathcal{S}, P_{YZ S |X})$ of input alphabet $\mc{X}$, two output alphabets $\mc{Y}$, $\mc{Z}$,  feedback alphabet $\mathcal{S}$ and a set of probability distributions  $P_{YZS |X}(\cdot|x)$ on $\mc{Y} \times \mc{Z} \times{S} $ for every $x \in \mc{X}$.  The channel satisfies the following conditions for all $n=1,2,\ldots$
\be
\begin{split}
& \text{Pr}( Y_n = y_n, Z_n=z_n, S_n  = s_n  | x^n , y^{n-1},  z^{n-1} , s^{n-1}) \\
& = P_{YZS|X} (y_n,z_n,s_n | x_n )
\end{split}
\ee
for all $(y_n, z_n, s_n) \in \mc{Y} \times \mc{Z} \times \mc{S}$  and $(x^n , y^{n-1},  z^{n-1} , s^{n-1}) \in \mc{Y}^{n-1} \times \mc{S}^{n-1} \times  \mc{Z}^{n-1}$.
The schematic is  shown in Figure \ref{fig:bc_fb}(b). We note that the broadcast channel with noiseless feedback from both receivers is a special case with
$\mc{S} = \mc{Y} \times \mc{Z}$, and $S_n=(Y_{n}, Z_{n})$.

\begin{defi}
An $(n,M_0,M_1,M_2)$ transmission system for a given broadcast channel with generalized feedback consists of
\begin{itemize}
\item A sequence of mappings for the encoder: for $m=1,2,\ldots,n$
\be
\begin{split}
e_m: \{1,2,\ldots,M_0\} \times \{1,2,\ldots,M_1\} & \\
\times \{1,2,\ldots,M_2\} \times \mc{S}^{m-1}  & \rightarrow \mc{X}.
\end{split}
\ee
\item A pair of decoder mappings:
\be
\begin{split}
& g_1: \mc{Y}^{n} \rightarrow \{1,2,\ldots,M_0\} \times \{1,2,\ldots,M_1\}, \\
& g_2: \mc{Z}^n \rightarrow \{1,2,\ldots,M_0\} \times \{1,2,\ldots,M_2\}.
\end{split}
\ee
\end{itemize}
\end{defi}
\emph{\textbf{Remark}}: Though we have defined the transmission system above for feedback delay $1$, all the results in this paper
hold for feedback with any finite delay $k$.

We use $W_0$ to denote the common message, and $W_1,W_2$ to denote the private messages of decoders $1$ and $2$, respectively. The messages $(W_0,W_1,W_2)$ are uniformly distributed over the set $\{1,2,\ldots,M_0\} \times \{1,2,\ldots,M_1\} \times \{1,2,\ldots,M_2\}$. The channel input at time $n$ is given by $X_n=e_n(W_0,W_1,W_2, S^{n-1})$. The average error probability of the above transmission system is given by
\be
\tau=\frac{1}{M_0M_1M_2} \sum_{k=1}^{M_0} \sum_{i=1}^{M_1} \sum_{j=1}^{M_2} P_{e}(i,j,k)
\ee
where $P_{e}(i,j,k)$ equals
\[ \text{Pr} \left( (g_1(Y^n),g_2(Z^n)) \neq ((k,i),(k,j)) | W_0,W_1,W_2 = k,i,j  \right). \]

\begin{defi} \label{def:ach_rates}
A triple  of non-negative real numbers $(R_0,R_1,R_2)$ is said to be achievable  for a given  broadcast channel with feedback if $\forall \epsilon>0$, there exists an $N(\epsilon)>0$ such that for all $n > N(\epsilon)$, there exists an $(n,M_0,M_1,M_2)$ transmission system satisfying the following constraints:
\be
\begin{split}
\frac{1}{n} \log M_0 \geq R_0 -\epsilon,  \quad &  \frac{1}{n} \log M_1  \geq R_1 -\epsilon, \\
\frac{1}{n} \log M_2  \geq R_2 -\epsilon, \quad & \tau \leq \epsilon.
\end{split}
\ee
The closure of the set of all achievable rate pairs is the capacity region of the channel.
\end{defi}

We now define the structure for the joint distribution of all the variables in our coding scheme.  Due to the block-Markov nature of the scheme, the random variables carrying the resolution information in each block depend on the variables corresponding to the previous block. In order to obtain a single-letter rate region, we need the random variables in each block to follow the same joint distribution, say $P$. Hence, after each block of transmission, we generate the variables for the next block using a Markov kernel $Q$ that has invariant distribution $P$. This will guarantee a stationary joint distribution $P$ in each block.

\begin{defi}
Given a broadcast channel  with  feedback $(\mc{X},\mc{Y},\mc{Z},\mc{S},P_{YZS|X})$,   define
$\mc{P}$ as the set of all distributions $P$ on $\mc{U} \times \mc{V}
\times \mc{A} \times \mc{B} \times \mc{C} \times \mc{X}
\times \mc{Y} \times \mc{Z} \times \mc{S}$ of the form
\[
P_{ABC} \  P_{UV|ABC} \ P_{X|ABCUV}\ P_{YZS|X},
\]
where $\mc{A}$, $\mc{B}$, $\mc{C}$, $\mc{U}$, and $\mc{V}$
are arbitrary sets. Consider two sets of random variables
$(U,V,A,B,C,X,Y,Z, S)$ and
$(\tilde{U},\tilde{V},\tilde{A}, \tilde{B},\tilde{C},\tilde{X},\tilde{Y},\tilde{Z}, \tilde{S})$ each having the
same distribution $P$.    For brevity, we often refer to the
collection   $({A},{B},S)$ as ${K}$, to $(\tilde{A},\tilde{B},\tilde{S})$ as $\tilde{K}$,
and to $\mc{A} \times \mc{B} \times \mc{S}$ as $\mc{K}$.
Hence
\[P_{\tilde{U} \tilde{V} \tilde{C} \tilde{K} \tilde{X} \tilde{Y} \tilde{Z}} = P_{UVCKXYZ}=P.\]

For a given $P \in \mc{P}$, define $\mc{Q}(P)$
as the set of conditional distributions $Q$
that satisfy the following consistency condition
\be
\begin{split}
& P_{ABC}(a,b,c)=\\
& \sum_{\stackrel{(\tilde{u},\tilde{v}, ,\tilde{k}, \tilde{c} ) \in}{\mc{U} \times \mc{V} \times \mc{K} \times \mc{C}} }
Q_{ABC|\tilde{U} \tilde{V} \tilde{K}\tilde{C}}
(a,b,c | \tilde{u},\tilde{v},\tilde{k},\tilde{c}) \
P_{UVKC}(\tilde{u},\tilde{v},\tilde{k},\tilde{c})
\end{split}
 \label{eq:cons_cond}
\ee
 for all $(a,b,c)$. Then for any $P \in \mc{P}$ and $Q \in \mc{Q}(P)$, the {joint}
 distribution of the two sets $(U,V,K,C,X, Y, Z)$ and
 $(\tilde{U},\tilde{V},\tilde{K},\tilde{C},\tilde{X}, \tilde{Y}, \tilde{Z})$ is
\be
P_{\tilde{U}\tilde{V} \tilde{K} \tilde{C} \tilde{X} \tilde{Y} \tilde{Z}}
\ Q_{ABC|\tilde{U} \tilde{V} \tilde{C} \tilde{K}}
\ P_{UVKXYZ|ABC}.
\ee
\label{def:joint_dist}
\end{defi}

With the above definitions, we have the following theorem.
\begin{thm} \label{thm:mainthm_a}
Given a broadcast channel with generalized feedback $(\mc{X},\mc{Y},\mc{Z},\mc{S},P_{YZS|X})$, for any
distribution $P \in \mc{P}$ and $Q \in \mc{Q}(P)$, the convex hull of the  following
region is achievable.
\begin{align}
& R_0+R_1  < I(\tilde{U}AC; Y\tilde{Y} | \tilde{A} \tilde{C}) - I(\tilde{V} \tilde{B} \tilde{S}; AC | \tilde{U}  \tilde{A}  \tilde{C} ) \\
& R_0+R_2 <  I(\tilde{V}BC;Z\tilde{Z} | \tilde{B} \tilde{C}) - I(\tilde{U} \tilde{A} \tilde{S}; BC | \tilde{V} \tilde{B} \tilde{C})  \\
& R_0 + R_1+ R_2 <  I(\tilde{U}AC; Y\tilde{Y} | \tilde{A} \tilde{C})  - I(\tilde{V} \tilde{B} \tilde{S}; AC | \tilde{U}  \tilde{A}  \tilde{C} ) \nonumber \\
& \qquad  \qquad \qquad  \quad + I(\tilde{V};C| \tilde{B} \tilde{C}) +  I(\tilde{V}B;Z\tilde{Z} | C \tilde{B} \tilde{C}) \nonumber \\
& \qquad \qquad  \qquad  \quad - I(\tilde{U} \tilde{A} \tilde{S} A; B | C  \tilde{V} \tilde{B} \tilde{C})  - \mc{T}  \\
& R_0 + R_1+ R_2 < I(\tilde{V}BC;Z\tilde{Z} | \tilde{B} \tilde{C}) - I(\tilde{U} \tilde{A} \tilde{S}; BC | \tilde{V} \tilde{B} \tilde{C}) \nonumber \\
& \qquad  \qquad \qquad  \quad + I(\tilde{U};C| \tilde{A} \tilde{C}) +  I(\tilde{U}A;Y\tilde{Y}|C\tilde{A}\tilde{C}) \nonumber\\
&   \qquad  \qquad \qquad  \quad - I(\tilde{V} \tilde{B} \tilde{S} B; A | C\tilde{U} \tilde{A} \tilde{C}) - \mc{T} \\
& 2R_0+R_1+R_2  < I(\tilde{U}AC; Y\tilde{Y} | \tilde{A} \tilde{C})  - I(\tilde{V} \tilde{B} \tilde{S}; AC | \tilde{U}  \tilde{A}  \tilde{C} )  \nonumber  \\
&\qquad  \qquad \qquad  \  \  + I(\tilde{V}BC;Z\tilde{Z} | \tilde{B} \tilde{C}) - I(\tilde{U} \tilde{A} \tilde{S}; BC | \tilde{V} \tilde{B} \tilde{C}) \nonumber \\
& \qquad  \qquad \qquad \  \  - I(A;B | C \tilde{C} \tilde{U} \tilde{V}  \tilde{K})  - \mc{T} \\
& R_0 < \min\{\mc{T}_1, \mc{T}_2, \mc{T}_3, \mc{T}_4, \mc{T}_5 \}
\end{align}
where $\tilde{K} = (\tilde{A}, \tilde{B}, \tilde{S})$ and
\begin{align*}
\mc{T} &\triangleq H(U|AC) + H(V|BC) - H(UV|ABC), \\
\mc{T}_1 &\triangleq I(AC; Y\tilde{Y}\tilde{A}|\tilde{C} \tilde{U})- I(\tilde{V} \tilde{K} ; AC |\tilde{C} \tilde{U}), \\
\mc{T}_2 &\triangleq I(BC; Z\tilde{Z}\tilde{B}|\tilde{C} \tilde{V})- I(\tilde{U} \tilde{K} ; BC |\tilde{C} \tilde{V}), \\
\mc{T}_3 &\triangleq  I(AC; Y\tilde{Y}\tilde{A}|\tilde{C} \tilde{U})  - I(\tilde{V} \tilde{K} ; AC |\tilde{C} \tilde{U}) \\
& \quad + I(B; Z\tilde{Z}\tilde{B}|\tilde{C} \tilde{V}C) - I(\tilde{U} \tilde{K}A ; B |C \tilde{C} \tilde{V}), \\
\mc{T}_4 &\triangleq  I(A; Y\tilde{Y}\tilde{A}|\tilde{C} \tilde{U}C)  - I(\tilde{U} \tilde{K} ; B C | \tilde{C} \tilde{V})  \\
&\quad + I(BC; Z\tilde{Z}\tilde{B}|\tilde{C} \tilde{V}) - I(\tilde{V} \tilde{K}B ; A |C \tilde{C} \tilde{U} ),  \\
\mc{T}_5 &\triangleq  \frac{1}{2} \left( \mc{T}_1 + \mc{T}_2 - I(A;B|C \tilde{C}\tilde{U}\tilde{V}\tilde{K}) \right).
\end{align*}
\end{thm}
\proof This theorem is proved in Section \ref{sec:proof}.

 \textbf{\emph{Remarks}}:
 \begin{enumerate}
 \item The input mapping $P_{X|ABCUV}$ in the set of distributions $\mc{P}$ can be assumed to be deterministic, i.e,  $X = f(A,B,C,U,V)$ for some function $f$. This is because for a fixed $P_{ABCUV}$, optimizing the rate region is equivalent to maximizing a convex functional of  $P_{X|ABCUV}$. Hence the optimum occurs at one of the corner points, which corresponds to a deterministic $P_{X|ABCUV}$.
\item  We can recover Marton's achievable rate region for the broadcast channel without feedback by  setting $A=B=\phi$, and $C=W$ with $Q_{C|\tilde{U} \tilde{V} \tilde{K}\tilde{C}} = P_W$.
\end{enumerate}

\section{Coding scheme} \label{sec:coding_scheme}
In this section, we give an informal outline of the proof of Theorem \ref{thm:mainthm_a}. The formal proof is given in Section \ref{sec:proof}. Let us first consider the case when there is no common message ($R_0 = 0$). Let the  message  rate pair $(R_1,R_2)$ lie outside Marton's achievable region \cite{Marton79}.
The coding scheme uses a block-Markov superposition strategy, with the communication taking place over $L$ blocks, each of length $n$.

In each block, a  fresh pair of messages is encoded using the Marton coding strategy (for the BC without feedback). In block $l$, random variables $U$ and $V$ carry the fresh information  for receivers $1$ and $2$, respectively. At the end of this block, the receivers are not able to decode the information in $(U,V)$ completely, so we send `resolution' information  in block $(l+1)$ using random variables $(A,B,C)$. The pair $(A,C)$ is meant to be decoded by the first receiver, and the pair $(B,C)$  by the second receiver. Thus in each block, we obtain the channel output by superimposing fresh information on  the resolution information for the previous block. At the end of the block, the first receiver decodes $(A,C)$,  the second receiver decodes $(B,C)$,  thereby resolving the uncertainty about their messages of the previous block.

\emph{Codebooks}: The $A$-, $B$-, and $C$-codebooks are constructed on the alphabets $\mc{A}$, $\mc{B}$, and $\mc{C}$ respectively.  The exact procedure for this construction, and the method for selecting codewords from these codebooks will be described in the sequel. Since $(A,C)$ is decoded first by receiver $1$, conditioned on each codeword pair corresponding to the $A$- and $C$-codebooks, we construct a $\mathbf{U}$-codebook of size $2^{nR'_1}$ by generating codewords according to $P_{U|AC}$. Similarly for each codeword pair in the $B$- and $C$-codebooks, we construct a $\mathbf{V}$-codebook of size $2^{nR'_2}$ by generating codewords according to $P_{V|BC}$. Each $\mathbf{U}$-codebook is divided into $2^{nR_1}$ bins, and each $\mathbf{V}$-codebook into $2^{nR_2}$ bins.

\emph{Encoding}: In each block $l$, the encoder chooses a tuple of five codewords $(\mathbf{A}_l,\mathbf{B}_l,\mathbf{C}_l, \mathbf{U}_l, \mathbf{V}_l)$ as follows. The resolution  information for block $(l-1)$ is used to select $(\mathbf{A}_l,\mathbf{B}_l,\mathbf{C}_l)$ from the $A$-, $B$- and $C$-codebooks. $\mathbf{C}_l$ determines the $\mathbf{U}$- and $\mathbf{V}$-codebooks to be used to encode the message pair  of block $l$.  Denoting the message pair by $(m_{1l},m_{2l})$, the encoder chooses a $U$-codeword from bin $m_{1l}$ of  the $U$-codebook and a $V$-codeword from bin $m_{2l}$ of the $V$-codebook that are jointly typical according to $P_{UV|ABC}$. This pair of jointly typical codewords is set to be $(\mathbf{U}_l,\mathbf{V}_l)$.

By standard joint-typicality based covering arguments (see e.g., \cite{ElGMuelen81}), this step is successful if the product of the sizes of $U$-bin and $V$-bin is exponentially larger than $2^{n(H(U|AC) + H(V|BC) - H(UV|ABC))}$ . Therefore, we have
\be
\label{eq:rate1}
R'_1+R'_2 -R_1 -R_2 > H(U|AC) + H(V|BC) - H(UV|ABC).
\ee
These five codewords are combined using the transformation $P_{X|ABCUV}$ (applied component-wise) to generate the channel input $\mathbf{X}_l$.

\emph{Decoding}: After receiving the channel output of block $l$, receiver $1$ first decodes $(\mathbf{A}_l,\mathbf{C}_l)$, and receiver $2$ decodes $(\mathbf{B}_l,\mathbf{C}_l)$. However, the rates $R'_1, R'_2$ of the $U$- and $V$-codebooks are too large for receivers $1$ and $2$ to uniquely decode $\mathbf{U}_l$ and
$\mathbf{V}_l$, respectively. Hence receiver $1$ is left with a list of ${U}$-codewords that are jointly typical with its channel output $\mathbf{Y}_{l}$ and the just-decoded resolution information $(\mathbf{A}_l,\mathbf{C}_l)$; receiver $2$ has a similar list of ${V}$-codewords that are jointly typical with its channel output $\mathbf{Z}_{l}$, and the just-decoded resolution information $(\mathbf{B}_l,\mathbf{C}_l)$. The sizes of the lists are nearly equal to $2^{n(R'_1-I(U;Y|AC))}$ and   $2^{n(R'_2-I(V;Z|BC))}$, respectively. The transmitter receives feedback signal $\mathbf{S}_l$ in block $l$, and resolves these lists in the next block as follows.

In block $(l+1)$, the random variables of block $l$ are represented using the notation $\sim$.
Thus we have
\begin{equation*}
\begin{split}
& \mathbf{\tilde U}_{l+1} = \mathbf{U}_l,\ \mathbf{\tilde V}_{l+1}= \mathbf{V}_l, \ \mathbf{\tilde C}_{l+1} = \mathbf{C}_{l},   \\
& \mathbf{\tilde A}_{l+1} = \mathbf{A}_{l},\ \mathbf{\tilde B}_{l+1} = \mathbf{B}_{l},\ \mathbf{\tilde S}_{l+1} = \mathbf{S}_{l}.
\end{split}
\end{equation*}
The random variables  $(U,V,A,B,C, Y, Z, S)$ in block  $l$ are jointly distributed via $P_{ABC}P_{UV|ABC}P_{YZS|ABCUV}$ chosen from $\mathcal{P}$ as given in
the statement of the theorem.

\begin{figure}
\centering
\includegraphics[width=3.4in]{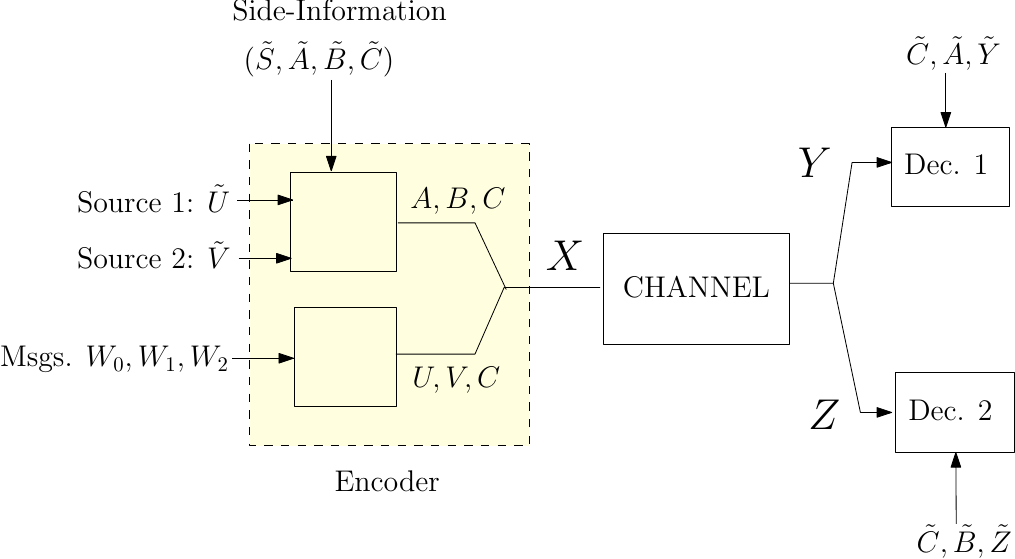}
\caption{\small{Transmitting correlated sources with side-information at the receivers through $(A,B,C)$, and fresh information through $U,V,C$. $C$ plays the dual role: it is used to cover the correlated sources and to carry fresh information.}}
\label{fig:corr_sources}
\vspace{-4pt}
\end{figure}

For block $l+1$,  $(\mathbf{\tilde U}_{l+1}, \mathbf{\tilde V}_{l+1})=(\mathbf{ U}_{l}, \mathbf{ V}_{l})$ can be considered to be a realization of a pair of correlated `sources'  ($\tilde{U}$ and $\tilde{V}$), jointly distributed according to $P_{\tilde{U}\tilde{V} \mid \tilde{S} \tilde{A}\tilde{B}\tilde{C}}$ along with the transmitter side information given by $(\mathbf{\tilde{A}}_{l+1},\mathbf{\tilde B}_{l+1}, \mathbf{\tilde S}_{l+1})$, and the common side-information $\mathbf{\tilde C}_{l+1}$. The goal in block $(l+1)$ is to transmit this pair of correlated  sources over the BC, with
\begin{itemize}
\item Receiver $1$ needing to decode $\mathbf{\tilde U}_{l+1}$, treating $(\mathbf{\tilde A}_{l+1},\mathbf{\tilde Y}_{l+1},\mathbf{\tilde C}_{l+1})$ as receiver side-information,
\item Receiver $2$ needing to decode $\mathbf{\tilde V}_{l+1}$,  treating $(\mathbf{\tilde{B}}_{l+1},\mathbf{\tilde Z}_{l+1},\mathbf{\tilde C}_{l+1})$ as receiver side-information.
\end{itemize}
We  use the ideas of Han and Costa \cite{HanCosta87} to transmit this pair of correlated sources over the BC (with appropriate extensions to take into account the different side-information available at the transmitter and the receivers).  This is shown in Figure \ref{fig:corr_sources}. The triplet of correlated random variables $(A,B,C)$ is used to cover the sources. This triplet carries  the resolution information intended to disambiguate the lists of the two receivers. The random variables
of block $(l+1)$, given by $(A,B,C)$ are related to the random variables in block $l$ via $Q_{ABC|\tilde{U}\tilde{V} \tilde{C} \tilde{A} \tilde{B}  \tilde{S}  }$, chosen from
$\mathcal{Q}$ given in the statement of the theorem. We now describe the construction of the $A$-, $B$-, and $C$- codebooks.

For brevity, we denote the collection of random variables $(A,B,S)$ as $K$, and $(\mathbf{A}_l,\mathbf{B}_l,\mathbf{S}_l)$ as
$\mathbf{K}_l=\mathbf{\tilde K}_{l+1}$.

\emph{Covering the Sources}:
For each $\mathbf{\tilde{c}} \in \mc{C}^n$, a $C$-codebook $\Psi_C(\mathbf{\tilde{c}})$ of rate $\rho_0$ is constructed randomly from $P_{C|\tilde{C}}$.
For every realization of $\mathbf{\tilde{u}} \in \mc{U}^n$, $\mathbf{\tilde{c}} \in \mc{C}^n$, and $\mathbf{c} \in \mathcal{C}^n$, an $A$-codebook $\Psi_A(\mathbf{\tilde{u}},\mathbf{\tilde{c}},\mathbf{c})$ of rate  $\rho_1$ is constructed with codewords picked  randomly according to $P_{A|\tilde{U} \tilde{C} C}$.
For every realization of $\mathbf{\tilde{v}} \in \mc{V}^n$, $\mathbf{\tilde{c}} \in \mc{C}^n$, and $\mathbf{c} \in \mathcal{C}^n$, a $B$-codebook
$\Psi_B(\mathbf{\tilde{v}},\mathbf{\tilde{c}},\mathbf{c})$ of rate  $\rho_2$ is constructed with codewords picked  randomly according to $P_{B|\tilde{V} \tilde{C} C}$.
 \footnote{We can also construct the $A$-codebook  with codewords picked  according to $P_{A |\tilde{U} \tilde{C} C \tilde{A}}$, and the $B$-codebook with codewords picked according  to $P_{B|\tilde{V} \tilde{C} C \tilde{B}}$.  Interestingly, this yields the same final rate region,  though the covering and packing conditions  are different.}

At the beginning of block $(l+1)$,
for a given  realization $(\mathbf{\tilde{U}}_{l+1},\mathbf{\tilde{V}}_{l+1},\tmb{K}_{l+1},\mathbf{\tilde{C}}_{l+1})$,
of correlated `sources',  and side information,
the encoder chooses a triplet of   codewords $(\mathbf{A}_{l+1},\mathbf{B}_{l+1},\mathbf{C}_{l+1})$ from the
appropriate $A$-, $B$- and $C$-codebooks such that the two tuples  are
jointly typical according to $P_{\tilde{U}\tilde{V}\tilde{K}\tilde{C}}
Q_{ABC|\tilde{U}\tilde{K}\tilde{V}\tilde{C}}$. The channel input $\mathbf{X}_{l+1}$ is generated by fusing this
$(\mathbf{A}_{l+1},\mathbf{B}_{l+1},\mathbf{C}_{l+1})$  with
the pair of codewords $(\mathbf{U}_{l+1},\mathbf{V}_{l+1})$,  which carry fresh information in block $(l+1)$.

Now consider the general case when $R_0>0$. We can use the random variable $C$ to encode common information to be decoded by both receivers. Hence $C$ serves two purposes: it is used to (a) cover the correlated sources and transmitter side-information and is thus part of the resolution information, and (b)  to carry fresh information that is decoded by both receivers. We note that  in every block, two communication tasks are being accomplished simultaneously. The first is joint source-channel coding of correlated sources over the BC, accomplished via $(A,B,C)$; the second is Marton coding of the fresh information, accomplished via $(U,V,C)$ \footnote{Recall that in Marton's achievable region for the  BC without feedback, there is a random variable $W$ meant to be decoded by both receivers.}.    $C$ can be made to assume the dual role of the common random variable associated with both these tasks.

\emph{Analysis}:
For this encoding to be successful, we need the following covering conditions. These are the same conditions that appear in the Han-Costa scheme (\cite[Lemma 14.1]{ElGKimBook}), with $(\tilde{U}, \tilde{K})$ and $(\tilde{V}, \tilde{K})$ assuming the roles of the two sources being covered.\footnote{Though $\tilde{K} = (\tilde{A}, \tilde{B}, \tilde{S})$ is included in the covering, it is not required to be explicitly decoded at either receiver. }
\begin{align}
& \rho_0 > I(\tilde{U} \tilde{K} \tilde{V};C|\tilde{C})+R_0  \label{eq:rate2} \\
& \rho_0+\rho_1 > I(\tilde{V}\tilde{K} ; A|C \tilde{C} \tilde{U}) +I(\tilde{U} \tilde{K} \tilde{V} ;C|\tilde{C}) +R_0 \\
& \rho_0+\rho_2 > I(\tilde{U} \tilde{K} ;B|C \tilde{C} \tilde{V}) +I(\tilde{U} \tilde{K} \tilde{V};C|\tilde{C}) +R_0 \\
& \rho_0+\rho_1+\rho_2  > I(\tilde{V}\tilde{K} ; A|C \tilde{C} \tilde{U})  + I(\tilde{U} \tilde{K} ;B|C \tilde{C} \tilde{V})  \label{eq:rate3} \\
& \qquad  \qquad +I(A;B| \tilde{U} \tilde{K} \tilde{V} C \tilde{C})+I(\tilde{U} \tilde{K} \tilde{V};C|\tilde{C})+R_0 \nonumber
\end{align}

At the end of block $(l+1)$, receiver $1$ determines $\mathbf{U}_{l}=\mathbf{\tilde{U}}_{l+1}$ by finding the  pair
$(\mathbf{\tilde{U}}_{l+1},\mathbf{A}_{l+1},\mathbf{C}_{l+1})$  using joint typical decoding in the composite $U$-, $A$-, and $C$-codebooks. A similar procedure is
followed at the second receiver. For decoding to be successful, we need the following packing conditions.
\begin{align}
& R'_1 + \rho_0 + \rho_1  < I(\tilde{U};Y\tilde{Y} | \tilde{A} \tilde{C}) + I(C;Y \tilde{A} \tilde{Y} \tilde{U}|\tilde{C}) \nonumber \\
& \qquad \qquad \qquad  \quad + I(A;Y\tilde{A}\tilde{Y} |\tilde{U}  \tilde{C} C)  \label{eq:rate4} \\
& R'_1+\rho_1 < I(\tilde{U};Y  \tilde{Y} C | \tilde{A} \tilde{C}) +I(A;Y\tilde{A} \tilde{Y} |\tilde{U} \tilde{C} C) \\
& R'_2 + \rho_0 + \rho_2 < I(\tilde{V};Z\tilde{Z} | \tilde{B} \tilde{C}) +I(C;Z \tilde{B} \tilde{Z} \tilde{V}|\tilde{C}) \nonumber \\
& \qquad \qquad \qquad  \quad + I(B;Z\tilde{B} \tilde{Z} |\tilde{V} \tilde{C} C)  \\
& R'_2+\rho_2 < I(\tilde{V};Z \tilde{Z} C| \tilde{B}  \tilde{C}) +I(B;Z\tilde{B} \tilde{Z} |\tilde{V} \tilde{C} C)  \\
& \rho_0+\rho_1 < I(C;Y\tilde{A} \tilde{Y}\tilde{U}|\tilde{C}) +I(A;Y\tilde{A}\tilde{Y}| \tilde{U} \tilde{C} C) \\
& \rho_0+\rho_2 < I(C;Z \tilde{B} \tilde{Z}\tilde{V}|\tilde{C}) +I(B;Z\tilde{B}\tilde{Z}| \tilde{V} \tilde{C} C)  \\
& \rho_1 < I(A;Y \tilde{A} \tilde{Y}| \tilde{U} \tilde{C} C) \\
& \rho_2 < I(B;Z \tilde{B} \tilde{Z}| \tilde{V} \tilde{C} C)
\label{eq:rate5}
\end{align}

Performing Fourier-Motzkin elimination on equations (\ref{eq:rate1}),
(\ref{eq:rate2}-\ref{eq:rate3}) and (\ref{eq:rate4}-\ref{eq:rate5}), we obtain the statement of the
theorem.

To get a single-letter characterization of achievable rates, we need to ensure
that the random variables in each block follow a stationary joint distribution.
We now ensure that the sequences in each block are jointly distributed according to
\be \label{eq:joint_dist}
P_{ABC}\cdot P_{UV|ABC} \cdot P_{X|ABCUV}\cdot P_{YZS|X}
\ee
for some chosen $P_{ABC}, P_{UV|ABC}$, and $P_{X|ABCUV}$.

Suppose that the sequences in a given block are jointly distributed according to \eqref{eq:joint_dist}. In the next block, these sequences become the source pair $(\tilde U,  \tilde V)$,  transmitter side-information $(\tilde{A},\tilde{B}, \tilde{C}, \tilde{S})$ and the side information at the two receivers -- $(\tilde{C}, \tilde{A}, \tilde{Y})$ and $(\tilde{C}, \tilde{B}, \tilde{Z})$, respectively. To cover the source pair with $(A,B,C)$, we  pick a conditional distribution $ Q_{ABC|\tilde A \tilde B \tilde C \tilde U  \tilde V \tilde S} $ such that the covering sequences are distributed according to $P_{ABC}$. This holds when the consistency condition given by \eqref{eq:cons_cond} is satisfied. We thereby ensure that the sequences in {each} block are jointly distributed according to \eqref{eq:joint_dist}. Our technique of exploiting the correlation  induced by feedback is similar in spirit to the coding scheme of Han for two-way channels \cite{Han84}.

We note that the transmitter side information $\tilde{K}=(\tilde{A}\tilde{B}\tilde{S})$ is exploited at the encoder in the covering operation implicitly, without using codebooks conditioned on $\tilde{K}$. This is because this side information is only partially available at the receivers, with receiver $1$ having only $(\tilde{A},\tilde{Y})$, and receiver $2$ having only $(\tilde{B},\tilde{Z})$.  Hence the coding approach does not depend on any assumptions on the nature of the generalized feedback signal $S$. This is in contrast to communication over a multiple-access channel with feedback, where there is a significant difference between  noiseless feedback and noisy feedback \cite{RamjiMAC11}.
\section{Special Cases and Examples} \label{sec:example}
We now obtain a simpler version of the region of Theorem \ref{thm:mainthm_a} and use it to compute achievable rates for
a few examples.
\subsection{A Simpler Rate Region}
\begin{corr} \label{corr:simpler_rate_region}
Given a broadcast channel  with generalized feedback $(\mc{X},\mc{Y},\mc{Z},\mc{S},P_{YZS|X})$,   define any joint distribution
$P$ of the form
\be P_{C_0} P_{WUV} P_{X|WUVC_0} P_{YZS|X}. \label{eq:sep_joint}\ee
for some discrete random variables  $W, U, V, C_0$.
Let $(C_0, W, U, V, X, Y, Z, S) $ and $(\tilde{C}_0, \tilde{W}, \tilde{U}, \tilde{V}, \tilde{X}, \tilde{Y}, \tilde{Z}, \tilde{S})$
be two sets of variables each distributed according to $P$ and jointly distributed as
\be
P_{\tilde{C}_0 \tilde{W} \tilde{U} \tilde{V} \tilde{X} \tilde{Y} \tilde{Z} \tilde{S}} \ Q_{C_0| \tilde{C}_0 \tilde{W} \tilde{U} \tilde{V} \tilde{S}}
\ P_{W U V X Y Z S|C_0}.
\ee
where $Q_{C_0|\tilde{C}_0 \tilde{W} \tilde{U} \tilde{V} \tilde{S}}$ is a distribution such that
\be
\begin{split}
 & P_{C_0}(c_0)= \\
 & \sum_{\tilde{c}_0,\tilde{w},\tilde{u}, \tilde{v},\tilde{s}} Q_{C_0|\tilde{C}_0 \tilde{W} \tilde{U} \tilde{V} \tilde{S}}
 (c_0|\tilde{c}_0, \tilde{w}, \tilde{u}, \tilde{v}, \tilde{s}) P(\tilde{c}_0, \tilde{w}, \tilde{u}, \tilde{v}, \tilde{s})
 \end{split}
\label{eq:c0_consist}
\ee
for all $c_0 \in \mc{C}_0$. Then the following region is achievable.
\begin{align}
& R_0   < \min\{\mc{T}_1, \mc{T}_2 \}\\
& R_0 + R_1  < I(UW;Y|C_0)+ I(C_0; Y|\tilde{Y} \tilde{C_0} \tilde{W}) \nonumber \\
& \qquad \qquad  + I(C_0 ; \tilde{Y}|\tilde{C}_0 \tilde{W} \tilde{U})  - I(\tilde{V} \tilde{S}; C_0 | \tilde{C}_0 \tilde{W} \tilde{U})  \\
& R_0 + R_2  < I(VW;Z|C_0)+ I(C_0; Z|\tilde{Z} \tilde{C_0} \tilde{W}) \nonumber \\
&\qquad \qquad  + I(C_0 ; \tilde{Z}|\tilde{C}_0 \tilde{W} \tilde{V}) - I(\tilde{U} \tilde{S}; C_0 | \tilde{C}_0 \tilde{W} \tilde{V})  \\
& R_0+R_1+R_2  < I(UW;Y|C_0)+ I(C_0; Y|\tilde{Y} \tilde{C_0} \tilde{W})  \nonumber \\
& \qquad \qquad  + I(C_0 ; \tilde{Y}|\tilde{C}_0 \tilde{W} \tilde{U}) - I(\tilde{V} \tilde{S}; C_0 | \tilde{C}_0 \tilde{W} \tilde{U})\\
&  \qquad \qquad + I(C_0\tilde{Z};\tilde{V}| \tilde{C}_0 \tilde{W})- I(U;V|W) \\
& R_0+R_1+R_2  < I(VW;Z|C_0)+ I(C_0; Z|\tilde{Z} \tilde{C_0} \tilde{W})  \nonumber \\
&\qquad \qquad+ I(C_0 ; \tilde{Z}|\tilde{C}_0 \tilde{W} \tilde{V}) - I(\tilde{U} \tilde{S}; C_0 | \tilde{C}_0 \tilde{W} \tilde{V}) \nonumber \\
& \qquad \qquad + I(C_0\tilde{Y};\tilde{U}| \tilde{C}_0 \tilde{W})-I(U;V|W)  \\
& 2R_0 + R_1 + R_2  < I(UW;Y|C_0)+  I(VW;Z|C_0) \nonumber \\
&\ + I(C_0; Y|\tilde{Y} \tilde{C_0} \tilde{W}) + I(C_0 ; \tilde{Y}|\tilde{C}_0 \tilde{W} \tilde{U}) - I(\tilde{V} \tilde{S}; C_0 | \tilde{C}_0 \tilde{W} \tilde{U}) \nonumber \\
& \ + I(C_0; Z|\tilde{Z} \tilde{C_0} \tilde{W}) + I(C_0 ; \tilde{Z}|\tilde{C}_0 \tilde{W} \tilde{V}) - I(\tilde{U} \tilde{S}; C_0 | \tilde{C}_0 \tilde{W} \tilde{V}) \nonumber\\
& \ -I(U;V|W)
\end{align}
where
\begin{align}
\mc{T}_1 & \triangleq  I(C_0;\tilde{Y}| \tilde{C}_0 \tilde{W} \tilde{U}) + I(C_0W; Y|\tilde{Y} \tilde{C_0} \tilde{W} \tilde{U})  \nonumber \\
& \quad - I(\tilde{V} \tilde{S}; C_0 | \tilde{C}_0 \tilde{W} \tilde{U}), \\
\mc{T}_2  & \triangleq  I(C_0;\tilde{Z}| \tilde{C}_0 \tilde{W} \tilde{V}) + I(C_0W; Z|\tilde{Z} \tilde{C_0} \tilde{W} \tilde{V})  \nonumber \\
& \quad - I(\tilde{U} \tilde{S}; C_0 | \tilde{C}_0 \tilde{W} \tilde{V}).
\end{align}
\end{corr}
\IEEEproof In Theorem \ref{thm:mainthm_a}, set $A=B=\phi$, and $C=(C_0, W)$, with
$ Q_{C|\tilde{C} \tilde{U} \tilde{V} \tilde{S}} =  Q_{C_0 W|\tilde{C_0} \tilde{W} \tilde{U}\tilde{V}\tilde{S}} = P_{W} Q_{C_0|\tilde{C_0} \tilde{W} \tilde{U}\tilde{V}\tilde{S}}. $
For this choice, we have $Q_{C|\tilde{C} \tilde{U} \tilde{V} \tilde{S}} \in \mc{Q}(P)$ if \eqref{eq:c0_consist} is satisfied.

\subsection{The AWGN  Broadcast Channel with Noisy Feedback} \label{subsec:awgn}
We  use Corollary \ref{corr:simpler_rate_region} to compute achievable rates for the scalar AWGN broadcast channel with noisy feedback from one receiver.
The obtained sum rate is compared with: a) the maximum sum rate in the absence of feedback, b) the achievable region of Bhaskaran \cite{Bhaskaran08} for the case of noiseless feedback from one receiver, and c) the Ozarow-Leung  achievable region \cite{OzarowLeung84} for noiseless feedback  from {both} receivers. We note that the coding schemes in both \cite{Bhaskaran08} and \cite{OzarowLeung84} are linear schemes based on Schalkwijk-Kailath coding for the AWGN channel \cite{SchalkwijkKailath66}, and cannot be used when there is noise in the feedback link \cite{KimLapWeiss}. Our rate region also includes the possibility of a common message to both receivers. The coding schemes of \cite{OzarowLeung84} and \cite{Bhaskaran08} are constructed only for private messages.

The channel, with $\mc{X}=\mc{Y}=\mc{Z}=\mathbb{R}$, is described by
\be \label{eq:awgnbc}
Y=X+N_1, \quad Z=X+N_2,
\ee
where $N_1,N_2$ are Gaussian noise variables (independent of the channel input $X$) with zero mean and covariance matrix
\[ K_{N_1,N_2} = \sigma^2 \begin{bmatrix} 1 & \rho \\ \rho & 1 \end{bmatrix} \]
where $\rho \in [-1,1]$.  The input sequence $\mathbf{x}$ for each block satisfies an average power constraint $\sum_{i=1}^n x_{i}^2 \leq nP$.
¤
In the absence of feedback, the capacity region of the AWGN broadcast channel is known \cite{Cover72,Bergmans74} and
can be obtained from Marton's inner bound using the following choice of random variables.
\[  V=\sqrt{\bar{\alpha} P} \; Q_2, \quad  U= \sqrt{\alpha P}\; Q_1 + \frac{\alpha P}{\alpha P + \sigma^2} V \]
where $\alpha \in (0,1)$, and  $Q_1, Q_2$ are independent $\mc{N}(0,1)$ random variables.
The Marton sum rate is then given by
\be  R_{\text{no-FB}} = I(V;Z) + I(U;Y) - I(U;V) = \frac{1}{2} \log \left(1 + \frac{P}{\sigma^2}\right). \label{eq:nofb_rate} \ee
This is essentially the `writing on dirty paper' coding strategy \cite{Costa84, GelfPin}: for the channel from $U$ to $Y$, $V$ can be considered as channel state information known at the encoder. We note that  an alternate way of achieving the no-feedback capacity region of the AWGN broadcast channel is through superposition coding \cite{Cover98}. \footnote{Theorem \ref{thm:mainthm_a} was established for a discrete memoryless broadcast channel with feedback. These theorems can be extended to the AWGN broadcast channel using a similar proof, recognizing that in the Gaussian case superposition is equivalent to addition. }

Using Corollary \ref{corr:simpler_rate_region}, we now compute an achievable region for the channel \eqref{eq:awgnbc} with noisy feedback from transmitter $1$ alone.The feedback signal is given by
\be
S=Y+N_f
\label{eq:noisyfb_def}
\ee
where $N_f$ is additive white Gaussian noise  on the feedback link distributed as $\mathcal{N}(0, \sigma_f^2)$. $N_f$ is independent of $X,Y,Z,N_1$ and $N_2$.
To motivate the choice of joint distribution, let us first consider the case of noiseless feedback, i.e., $N_f=0$.

\emph{Noiseless Feedback}: The joint distribution $P_{C_0} P_{UV} P_{X|C_0 UV}$ is chosen as
\begin{gather}
\quad V = \sqrt{\bar{\alpha} P_1} \; Q_2, \quad  U = \sqrt{\alpha P_1} \; Q_1 + \beta \; V \\
X=\sqrt{P-P_1}\; C_0 +  \sqrt{\bar{\alpha} P_1}\; Q_2 + \sqrt{\alpha P_1} \; Q_1
\label{eq:x_inp_def}
\end{gather}
where $Q_1, Q_2, C_0$ are independent Gaussians with zero mean and unit variance and  $\alpha, \beta \in (0,1)$, $P_1 \in (0,P)$ are
parameters to be optimized later.

Next we define a conditional distribution $Q_{C_0|\tilde{C}_0 \tilde{U} \tilde{V} \tilde{Y} \tilde{Z}}$ that satisfies \eqref{eq:c0_consist}. Let
\be
\tilde{T}_1 = \frac{\tilde{U} - E[\tilde{U} | \tilde{Y} \tilde{C}_0]}{\sqrt{E[(\tilde{U} - E[\tilde{U}| \tilde{Y} \tilde{C}_0])^2]}}.
\label{eq:gen_T1til}
\ee
Then define $Q_{C_0|\tilde{C}_0 \tilde{U} \tilde{V} \tilde{Y} \tilde{Z}}$ by the relation
\be
C_0= \sqrt{1-D} \ \tilde{T}_1 + \zeta
\label{eq:C0_def}
\ee
where $\zeta$ is a $\mc{N}(0,D)$ random variable  independent of $(\tilde{C_0}, \tilde{U}, \tilde{V}, \tilde{Y}, \tilde{Z})$.

In words, $\tilde{T}_1$ is the normalized error in the estimate of $\tilde{U}$ at receiver $1$. This estimation error is quantized at distortion level $D$ and suitably scaled to obtain $C_0$. Thus, in each block, $C_0$ represents a quantized version of the estimation error at receiver $1$ in the previous block.
If we similarly denote  by $\tilde{T}_2$ the error in the estimate of $\tilde{V}$ at receiver $2$ (replacing $\tilde{U}, \tilde{Y}$ in \eqref{eq:gen_T1til} with $\tilde{V}, \tilde{Z}$), then $\tilde{T}_2$ is correlated with $\tilde{T}_1$. This can be seen by expressing  the  estimation errors as
\begin{align}
& \tilde{U} - E[\tilde{U} | \tilde{Y} \tilde{C}_0]  =  \sqrt{\alpha P_1} \left( \frac{\sigma^2 +  \bar{\alpha} \bar{\beta} P_1}{P_1 +\sigma^2} \right) \tilde{Q}_1
\nonumber \\
& \qquad \qquad + \sqrt{\bar{\alpha}P_1} \left( \frac{\beta \sigma^2 -  {\alpha} \bar{\beta}P_1}{P_1 +\sigma^2} \right) \tilde{Q}_2
- \frac{P_1(\alpha + \beta \bar{\alpha})}{P_1+\sigma^2} \tilde{N}_1,  \label{eq:est_err1}\\
& \tilde{V} - E[\tilde{V} | \tilde{Z} \tilde{C}_0]    =  - \sqrt{\alpha P_1} \frac{\bar{\alpha} P_1}{P_1 +\sigma^2} \ \tilde{Q}_1  \nonumber \\
& \qquad \qquad \qquad \quad + \sqrt{\bar{\alpha}P_1} \frac{\alpha P_1}{P_1 +\sigma^2} \ \tilde{Q}_2  - \frac{\bar{\alpha}P_1}{P_1 + \sigma^2} \tilde{N}_2.
\label{eq:est_err2}
\end{align}

 We see that correlation coefficient between the estimation errors in \eqref{eq:est_err1} and \eqref{eq:est_err2} depends on $\alpha, \beta, \rho$. As long as the correlation is non-zero, $C_0$ simultaneously plays the role of conveying information about $\tilde{T}_1$ to receiver $1$, and about $\tilde{T}_2$ to receiver $2$.
\begin{figure*}[t]
\centering
\includegraphics[width=5.8in]{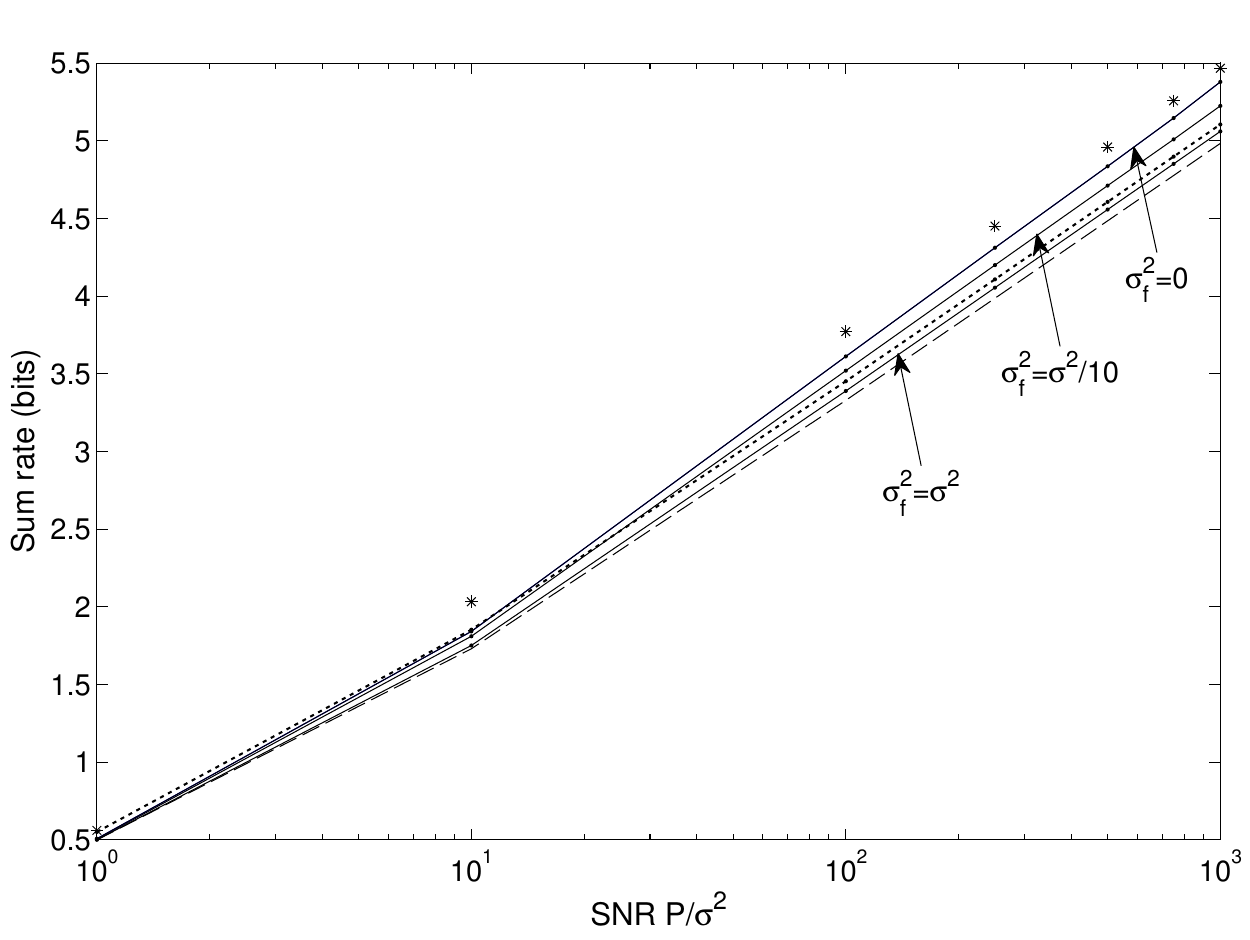}
\caption{\small{Achievable sum rates for the AWGN broadcast channel with noisy feedback from receiver $1$. Noise correlation $\rho=0$. The three solid lines show the sum-rates computed using Corollary \ref{corr:simpler_rate_region} for feedback noise variance $\sigma_f^2 = 0,; \sigma^2/10$, and  $\sigma^2$. The dashed line at the bottom is the no-feedback sum rate, the dotted line in the middle is the sum-rate of the Bhaskaran scheme, and the $\ast$ symbols at the top are the sum rate of the Ozarow-Leung scheme.}}
\vspace{-3pt}
\label{fig:awgn_plot}
\end{figure*}
\begin{figure}
\centering
\includegraphics[width=3.4in]{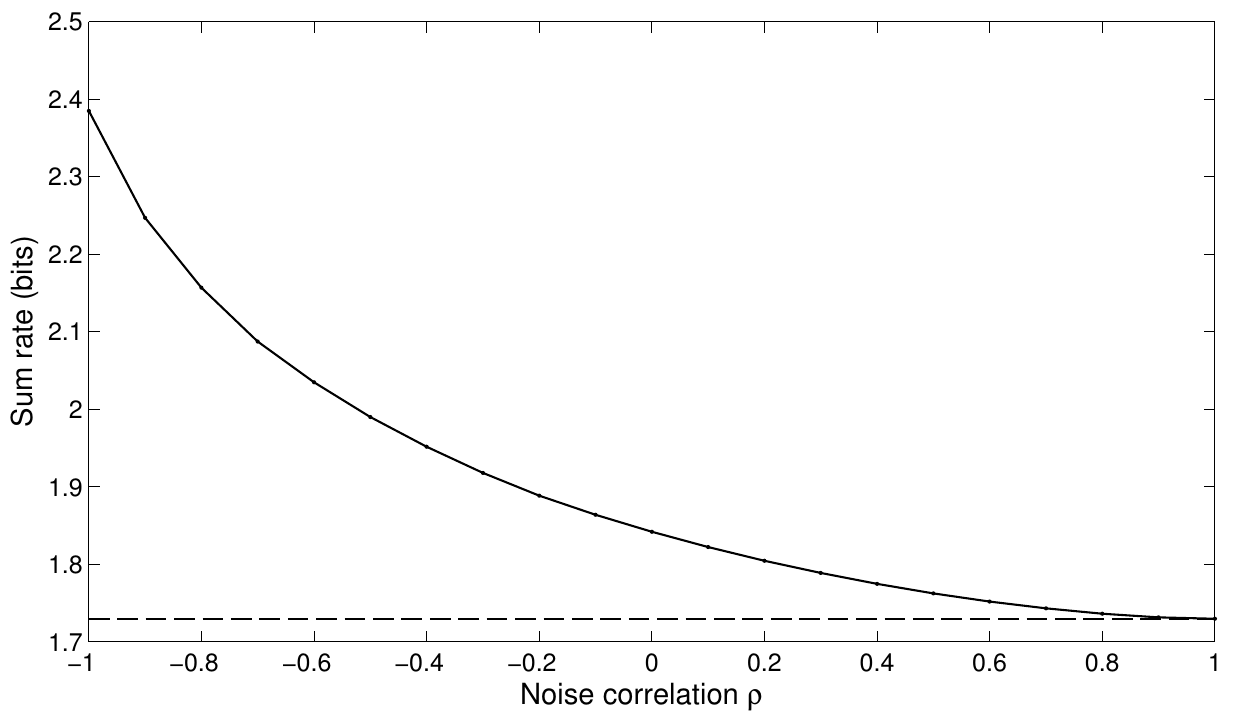}
\caption{\small{Variation of the sum-rate vs correlation coefficient of $(N_1,N_2)$. $P/\sigma^2=10$ and there is noiseless feedback from receiver $1$. The dashed line shows the no-feedback sum-rate}.}
\label{fig:corr_plot}
\vspace{-3pt}
\end{figure}
With this choice of joint distribution, the information quantities in Corollary \ref{corr:simpler_rate_region}
can be computed.

\emph{Noisy Feedback}: When the feedback is noisy, the transmitter does not know $\tilde{Y}$, and so cannot compute $\tilde{U} - E[\tilde{U}| \tilde{Y} \tilde{C}_0]$ in \eqref{eq:gen_T1til} which was used to generate $C_0$. Instead, the transmitter can compute an \emph{estimate} of the error at receiver $1$. We now define $\tilde{T}_1$ as
\be
\tilde{T}_1 = \frac{\Delta}{\sqrt{E[\Delta^2]}} 
\label{eq:gen_T1til_noisy}
\ee
where
\be
\begin{split}
\Delta & = E\left[ (\tilde{U} - E[\tilde{U} | \tilde{Y} \tilde{C}_0]) \mid \tilde{U}\tilde{V}\tilde{C_0} \tilde{S}  \right] \\
& = \sqrt{\alpha P_1} \frac{\sigma^2 +  \bar{\alpha} \bar{\beta} P_1}{P_1 +\sigma^2} \ \tilde{Q}_1 \
+ \sqrt{\bar{\alpha}P_1} \frac{\beta \sigma^2 -  {\alpha} \bar{\beta}P_1}{P_1 +\sigma^2} \ \tilde{Q}_2 \\
& \quad - \frac{P_1(\alpha + \beta \bar{\alpha})}{P_1+\sigma^2} \frac{\sigma^2}{\sigma^2 + \sigma_f^2} (\tilde{S} - \tilde{X}).
\end{split}
\ee
As before, $C_0$ is defined by \eqref{eq:C0_def} with $\tilde{T}_1$ given by \eqref{eq:gen_T1til_noisy}, and the input $X$ is defined by \eqref{eq:x_inp_def}.
With this choice of joint distribution, the information quantities required to evaluate Corollary \ref{corr:simpler_rate_region} are computed and
listed in Appendix \ref{app:awgn_compute}.

For different values of the signal-to-noise ratio $P/\sigma^2$, feedback noise variance $\sigma_f^2$ and  correlation coefficient $\rho$, we can compute the maximum sum rate by numerically optimizing over the parameters $(\alpha, \beta, D, P_1)$. For the case where the noises at the two receivers are independent ($\rho=0$), the maximum sum rate is plotted in Figure \ref{fig:awgn_plot} for  $\sigma_f^2=\sigma^2, \frac{\sigma^2}{10}$ and $0$ ($\sigma_f^2=0$ is noiseless feedback). The figure also shows the sum rate in the absence of feedback, the sum rate of the Bhaskaran scheme \cite{Bhaskaran08} for noiseless feedback from one receiver, and the maximum sum rate of the Ozarow-Leung scheme with noiseless feedback from {both} receivers.

We see that the obtained sum rate is higher than the no-feedback sum rate even with feedback noise variance $\sigma_f^2=\sigma^2$, and increases as $\sigma_f^2$ decreases. We also observe that for $\sigma_f^2=0$ (noiseless feedback), the sum rate of the proposed rate-region is higher than the Bhaskaran sum rate for high SNR. Concretely, for $P/\sigma^2 = 10, 100 $ and $1000$, our region yields sum rates of $1.842, 3.612$ and $5.378$, respectively; the Bhaskaran sum rates  for these SNR values are $1.852, 3.452$ and $5.105$. The Ozarow-Leung scheme yields higher sum rates than the proposed region. It also has the advantages of being a deterministic scheme with probability of error decaying double exponentially (with block length), but we emphasize that it uses noiseless feedback from both receivers. Another difference is that both the Ozarow-Leung and Bhaskaran schemes are specific to the AWGN broadcast channel and do not extend to other discrete memoryless broadcast channels, unlike the scheme in this paper.

Figure \ref{fig:corr_plot} shows the effect of $\rho$ (the correlation coefficient of $N_1, N_2$) on the sum-rate with $P/\sigma^2$ held fixed. The sum-rate without feedback does not change with $\rho$ as long as the individual noise variances remain unchanged \cite{Cover98}.
With noiseless feedback from receiver $1$,  the sum-rate obtained above decreases monotonically with the noise correlation and is equal to the no-feedback rate at $\rho=1$. This is consistent with the fact that feedback does not increase the capacity of  the AWGN broadcast channel with $\rho=1$ since it is physically degraded (in fact, we effectively have a point-to-point channel when $\rho=1$).
\subsection{Comparison with the Shayevitz-Wigger (S-W) Rate Region}
\label{sec:compare}
An achievable rate region for the broadcast channel with feedback was independently proposed by Shayevitz and Wigger \cite{WiggerShay}. Their coding scheme can be summarized as follows. In the first block, the encoder transmits at rates outside than the Marton region. The receivers cannot decode,   and as discussed earlier, the information needed to resolve the ambiguity at the two receivers is correlated. This resolution information is transmitted in the next block through separate source  and channel coding. The correlated resolution information is first quantized into three parts: a common part, and a private part for each receiver. This quantization is performed using a generalization of Gray-Wyner coding \cite{GrayWyner}. The quantization indices representing the correlated information  are then transmitted together with fresh information for the second block using Marton coding.

While the S-W scheme is also a block-Markov superposition scheme with the Marton coding as the starting point, the S-W scheme differs from the one proposed in this paper in two aspects:
\begin{enumerate}
\item Separate source and channel coding

\item Backward decoding
\end{enumerate}

While separate source and channel coding can be considered  a special case of joint source-channel coding, the backward decoding technique in \cite{WiggerShay} uses the resolution information in a different way than our scheme. In particular, the covering random variables in each block are decoded first and serve as extra `outputs' at the receivers that augment the channel  outputs. This difference in the decoding strategy makes a general comparison of the two rate regions difficult.

In Appendix \ref{app:comparison}, we show that the class of  valid joint distributions for the S-W region  can be obtained using our coding scheme via a specific choice of the covering variables $(A,B,C)$. The rate region of Theorem \ref{thm:mainthm_a} evaluated with this class of distributions is given in \eqref{eq:sw_eval1}--\eqref{eq:sw_eval6}. We observe that the bounds on $R_0 + R_1$, $R_0+R_2$, $R_0 + R_1 + R_2$ and $2R_0 + R_1 + R_2$  are larger than the corresponding bounds in the S-W region. However, our region has an additional $R_0$ constraint which is not subsumed by the other constraints. Therefore a general statement about the inclusion of  one region in the other does not seem  possible. In the following, we focus on the two examples discussed in \cite{WiggerShay} and show that the  feedback rates  of the S-W region can also be obtained using Corollary \ref{corr:simpler_rate_region}.

\emph{The Generalized Dueck Broadcast Channel}: This is a generalization of the Dueck example discussed in Section \ref{sec:intro}. The input $X$ is a binary triple $(X_0, X_, X_2)$. The output of the two receivers $1$ are $Y =(X_0 + N_0, X_1+ N_1)$ and $Z =(X_0 + N_0, X_1+ N_2)$ where $(N_0,N_1,N_2)$ are binary random variables with distribution $P_{N_0, N_1,N_2}$ such that
\[ H(N_0,N_1) \leq 1, \quad H(N_0,N_2) \leq 1.\]
We evaluate the rate region of Corollary \ref{corr:simpler_rate_region} for noiseless feedback from receiver $1$ with the following joint distribution.
\be
\begin{split}
(W,U,V) \sim &  P_W P_U P_V  \text{ with } P_W, P_U, P_V \sim \text{ Bernoulli}\left(\tfrac{1}{2}\right), \\
& Q_{C_0|\tilde{C_0} \tilde{W} \tilde{U} \tilde{V} \tilde{Y}}: C_0 = \tilde{Y} \oplus \tilde{U} =  \tilde{N}_1, \\
& X: (X_0, X_1, X_2) = (W, U, V)
\end{split}
\ee
With this choice of $Q$, $C_0$ is a Bernoulli random variable with the same distribution as $N_1$. With the joint distribution above, the mutual information quantities  in Corollary \ref{corr:simpler_rate_region} can be computed to be
\ben
\begin{split}
& I(U W;Y|C_0)=  2-H(N_0,N_1),  \\
& I( V W ; Z | C_0)= 2-H(N_0,N_2),  \\
& I(C_0; Y|\tilde{Y} \tilde{C_0} \tilde{W})= I(C_0; Z|\tilde{Z} \tilde{C_0} \tilde{W}) = 0, \\
& I(C_0 ; \tilde{Y}|\tilde{C}_0 \tilde{W} \tilde{U}) = H(N_1),  \\
& I(C_0 ; \tilde{Z}|\tilde{C}_0 \tilde{W} \tilde{V}) = H(N_1) - H(N_1|N_0, N_2), \\
& I(C_0 \tilde{Y}; \tilde{U}| \tilde{C}_0 \tilde{W}) = 1, \\
& I(C_0 \tilde{Z}; \tilde{V}| \tilde{C}_0 \tilde{W}) = 1 - H(N_2 | N_0 N_1),  \\
& I(\tilde{V} \tilde{S}; C_0 | \tilde{C}_0  \tilde{W}  \tilde{U})= I(\tilde{U} \tilde{S}; C_0 | \tilde{C}_0  \tilde{W}  \tilde{V}) = H(N_1), \\
& I(C_0W ; Y|\tilde{Y} \tilde{C}_0 \tilde{W} \tilde{U}) = I(C_0W ; Z|\tilde{Z} \tilde{C}_0 \tilde{W} \tilde{Z}) = 1- H(N_0).
\end{split}
\een
The rate region is given by
\be
\begin{split}
& R_0   \leq 1- H(N_0) - H(N_1|N_0, N_2) \\
& R_0 + R_1 \leq  2 - H(N_0, N_1) \\
& R_0 + R_2  \leq  2- H(N_0, N_1, N_2) \\
& R_0 + R_1 + R_2  \leq 3 - H(N_0, N_1, N_2)
\end{split}
\label{eq:region1}
\ee
 The roles of $R_1, R_2$ in \eqref{eq:region1} can be exchanged by choosing $C_0 = \tilde{Z} \oplus \tilde{V} =  \tilde{N}_2$.
Thus the following feedback capacity region obtained in \cite{WiggerShay} is achievable.
\be \begin{split} & R_1 \leq  2 - H(N_0, N_1), \quad R_2 \leq  2 - H(N_0, N_2), \\
& R_1 + R_2 \leq  3 - H(N_0, N_1, N_2). \end{split} \ee

\emph{The Noisy Blackwell Broadcast Channel}: This generalization of the Blackwell channel has ternary input alphabet $\mc{X} = \{0,1,2\}$, binary output alphabets $\mc{Y} = \mc{Z} = \{0,1\}$ and channel law given by
\ben
\begin{split}
Y= \left\{
\begin{array}{ll}
 N & X=0 \\
 1-N & X=1,2
\end{array}
\right.
\qquad
Z= \left\{
\begin{array}{ll}
 N & X=0,1 \\
 1-N & X=2
\end{array}
\right.
\end{split}
\een
where $N \sim \text{Bernoulli}(p)$ is a noise variable independent of $X$.
With noiseless feedback from both receivers, the rate region obtained in  \cite{WiggerShay} can also be obtained using Corollary \ref{corr:simpler_rate_region} with the following joint distribution.
\ben
\begin{split}
&P_W(0) =P_W(1) = \tfrac{1}{2}, \\
&P_{UV|W}(0,0|W=0)= \alpha, \ P_{UV|W}(1,1|W=0)= \beta, \\
& P_{UV|W}(1,0|W=0)= 1- \alpha - \beta, \\
&P_{UV|W}(0,0|W=1)= \beta, \ P_{UV|W}(1,1|W=1)= \alpha, \\
& P_{UV|W}(1,0|W=1)= 1- \alpha - \beta, \\
&X=U+V, \quad  Q_{C_0|\tilde{C_0} \tilde{W} \tilde{U} \tilde{V} \tilde{Y}}: C_0 = \tilde{Y} \oplus \tilde{U} = \tilde{Z} \oplus \tilde{V} = \tilde{N}.
\end{split}
\een
With $h(.)$ denoting the binary entropy function and $x\star y = x(1-y) + y (1-x)$, the mutual information quantities in Corollary \ref{corr:simpler_rate_region} are
\ben
\begin{split}
& I(U W;Y|C_0)=   I( V W ; Z | C_0)= h\left( p \star \frac{\alpha + \beta}{2}\right) - h(p), \\
& I(C_0; Y|\tilde{Y} \tilde{C_0} \tilde{W})= I(C_0; Z|\tilde{Z} \tilde{C_0} \tilde{W}) = 0, \\
&  I(\tilde{V} \tilde{S}; C_0 | \tilde{C}_0  \tilde{W}  \tilde{U})= I(\tilde{U} \tilde{S}; C_0 | \tilde{C}_0  \tilde{W}  \tilde{V}) = h(p),\\
&  I(C_0 \tilde{Z}; \tilde{V}| \tilde{C}_0 \tilde{W}) - I(U;V|W), \\
&  = H(V|UW) = \frac{1}{2}\left(\bar{\beta}
h\left( \frac{\alpha}{\bar{\beta}}\right) + \bar{\alpha}h\left(\frac{\beta}{\bar{\alpha}}\right) \right), \\
&I(C_0 \tilde{Y}; \tilde{U}| \tilde{C}_0 \tilde{W}) - I(U;V|W) \\
&  = H(U|VW) = \frac{1}{2}\left(\bar{\beta},
h\left( \frac{\alpha}{\bar{\beta}}\right) + \bar{\alpha}h\left(\frac{\beta}{\bar{\alpha}}\right) \right),\\
& I(W ; Y|\tilde{Y} \tilde{C}_0 \tilde{W} \tilde{U}) = I(W ; Z|\tilde{Z} \tilde{C}_0 \tilde{W} \tilde{Z}), \\
& = h\left( p \star \frac{\alpha + \beta}{2}\right) - \frac{1}{2} h\left(\frac{\alpha p + \bar{\alpha}\bar{p}}{2} \right)
- \frac{1}{2}\left( \frac{\beta p + \bar{\beta}\bar{p}}{2} \right).
\end{split}
\een
The rate region is then given by
\be
\begin{split}
& R_0  \leq h\left( p \star \frac{\alpha + \beta}{2}\right) - \frac{1}{2} h\left(\frac{\alpha p + \bar{\alpha}\bar{p}}{2} \right)
- \frac{1}{2} h\left( \frac{\beta p + \bar{\beta}\bar{p}}{2} \right) \\
& R_0 + R_1 \leq  h\left( p \star \frac{\alpha + \beta}{2}\right) - h(p) \\
& R_0 + R_2  \leq  h\left( p \star \frac{\alpha + \beta}{2}\right) - h(p) \\
& R_0 + R_1 + R_2  \leq h\left( p \star \frac{\alpha + \beta}{2}\right) - h(p) \\
& \qquad \qquad \qquad \qquad + \frac{1}{2}\left(\bar{\beta}
h\left( \frac{\alpha}{\bar{\beta}}\right) + \bar{\alpha}h\left(\frac{\beta}{\bar{\alpha}}\right) \right)
\end{split}
\label{eq:noisy_blackwell}
\ee
For $R_0 =0$, this matches the rate-region obtained in \cite{WiggerShay} for this channel.
\section{Proof  of Theorem \ref{thm:mainthm_a}}\label{sec:proof}
\subsection{Preliminaries} \label{subsec:prelim}
We shall use the notion of typicality as defined in \cite{OrlitskyRoche, ElGKimBook}.
Consider finite sets $\mathcal{Z}_1, \mathcal{Z}_2$ and any distribution $P_{Z_1Z_2}$ on them.
\begin{defi}
For any $\e>0$, the set of jointly $\e$-typical sequences with respect to $P_{Z_1Z_2}$ is defined as
\ben
\begin{split}
 \mc{A}^{(n)}_{\e}(P_{Z_1 Z_2}) = & \left\{ (\mathbf{z}_1, \mathbf{z}_2):  \left| \frac{1}{n} {N}(a,b \mid \mathbf{z}_1, \mathbf{z}_2) -P_{Z_1Z_2}(a,b) \right| \right. \\
& \  \leq \epsilon P_{Z_1Z_2}(a,b), \ \text{ for all } (a,b) \in \mc{Z}_1 \times \mc{Z}_2 \Big\}
\end{split}
\een
\end{defi}
where ${N}(a,b \mid \mathbf{z}_1, \mathbf{z}_2)$ is the number of occurrences of the symbol pair $(a,b)$ in the sequence pair $(\mathbf{z}_1, \mathbf{z}_2)$.
For any $\mathbf{z}_1 \in \mc{Z}_1^n$, define the set of conditionally $\e$-typical sequences as
\[ \mc{A}_{\e}^n(Z_2|\mathbf{z}_1) = \left\{ \mathbf{z}_2 : (\mathbf{z}_1, \mathbf{z}_2) \in \mc{A}^n_{\e}(P_{Z_1 Z_2}) \right\}. \]

The following are some basic properties of typical sequences that will be used in the proof. $\delta(\e)$ will be used to denote a generic positive function of $\e$ that tends to zero as $\e \to 0$.

\noindent \textbf{Property 0:} For all $\epsilon>0$, and for all
sufficiently large $n$, we have $P_{Z_1,Z_2}^n[\mc{A}^{(n)}_{\epsilon}(P_{Z_1, Z_2})] > 1-\epsilon$.

\noindent \textbf{Property 1:} Let $\mathbf{z}_1 \in \mc{A}^{(n)}_{\epsilon}(P_{Z_1})$ for some $\epsilon>0$. If  $\mathbf{Z}_2$ is generated according to the product distribution $\prod_{i=1}^n P_{Z_2|Z_1}(\cdot|z_{1i})$, then  for all $\e' >\e$
\[ \lim_{n \to \infty}\text{Pr}[( \mathbf{z}_1, \mathbf{Z}_2)  \in \mc{A}^{(n)}_{\e'}(P_{Z_1Z_2})] =1. \]

\noindent \textbf{Property 2:} For every $\mathbf{z}_1 \in \mc{Z}_1^n$, the size of the conditionally $\e$-typical set is upper bounded as
\[ | \mc{A}_{\e}^n(Z_2 | \mathbf{z}_1) | \leq 2^{n (H(Z_2|Z_1) + \delta(\e))}. \]
If $\mathbf{z}_1 \in \mc{A}_{\e}^n(P_{Z_1})$, then for any $\e' > \e$ and $n$ sufficiently large
\[ | \mc{A}_{\e}^n(Z_2 | \mathbf{z}_1) | \geq 2^{n (H(Z_2|Z_1) - \delta(\e'))}. \]

\noindent \textbf{Property 3:} If $(\mathbf{z}_1, \mathbf{z}_2) \in \mc{A}_{\e}^n(P_{Z_1,Z_2})$, then
\[ 2^{-n(H(Z_2|Z_1) + \delta(\e))} \leq P_{Z_2|Z_1}(\mathbf{z}_2 | \mathbf{z}_1) \leq 2^{-n(H(Z_2|Z_1) - \delta(\e))}.    \]

The definitions and properties above can be generalized in the natural way to tuples of multiple random variables as well.
\subsection{Random Codebook Generation}
We recall that $K$ denotes the collection $(A,B,S)$, and $\mc{K}$ denotes the set $\mc{A} \times \mc{B} \times \mc{S}$.

Fix a distribution $P_{UVABCXYZS}$ from $\mc{P}$ and a conditional distribution $Q_{ABC|\tilde{U}\tilde{V}\tilde{K}\tilde{C}}$ satisfying \eqref{eq:cons_cond},
as required by  the statement of the theorem. Fix a positive integer $L$. There are $L$ blocks in encoding
and decoding. Fix positive real numbers $R'_1,R'_2$, $R_0$,$R_1,R_2$, $\rho_0$, $\rho_1$ and $\rho_2$ such that $R'_1>R_1$ and
$R'_2>R_2$, where these numbers denote the rates of codebooks to be constructed as described below.
Fix block length $n$  and $\e>0$.  Let ${\e_l, \ l=1, \ldots,L}$ be  numbers such that
$\e < \e_1 < \e_2< \ldots <\e_L$.

\noindent For $l=1,2,3,\ldots,L$ independently perform the following random
experiments.
\begin{itemize}
\item For each sequence $\tilde{\mathbf{c}} \in \mc{C}^n$, generate   $2^{n \rho_0}$ sequences $\mathbf{C}_{[l,i,\tilde{\mathbf{c}}]}$,
  $i=1,2,\ldots,2^{n \rho_0}$, independently where each sequence is   generated from the product distribution $\prod_{i=1}^n P_{C|\tilde{C}} (\cdot|\tilde{c}_i)$.

\item For each sequence pair $(\mathbf{c}, \mathbf{a}) \in \mc{C}^n \times \mc{A}^n$, generate   $2^{n(R'_1-R_1)}$ sequences $\mathbf{U}_{[l,i,\mathbf{c}, \mathbf{a}]}$,   $i=1,2,\ldots,2^{n(R'_1-R_1)}$, independently where each sequence is generated from the product distribution $\prod_{i=1}^n P_{U|AC}(\cdot|a_i, c_i)$. Call this the first $U$-bin. Independently repeat this experiment  $2^{nR_1}$ times to generate
  $2^{nR_1}$ $U$-bins, and a total of $2^{nR'_1}$ sequences. The $i$th sequence in the $j$th bin is $\mathbf{U}_{[l,(j-1)2^{nR_1}+i, \: \mathbf{c}, \mathbf{a} ]}$.

\item For each sequence pair $(\mathbf{c}, \mathbf{b}) \in \mc{C}^n \times \mc{B}^n$, similarly   generate $2^{nR_2}$ $V$-bins each containing
  $2^{n(R'_2-R_2)}$ sequences with each sequence being generated from  the product distribution $\prod_{i=1}^n P_{V|BC}(\cdot|b_i, c_i)$. The $i$th sequence in the $j$th bin is $\mathbf{V}_{[l,(j-1)2^{nR_2}+i, \: \mathbf{c}, \mathbf{b} ]}$.

\item For each $(\tmb{u},\tmb{c},\mathbf{c}) \in \mc{U}^n \times \mc{C}^{n} \times \mc{C}^{n}$ generate independently   $2^{n\rho_1}$ sequences $\mathbf{A}_{[l,i,\tmb{u},\tmb{c},\mathbf{c}]}$,  for $i=1,2,\ldots,2^{n \rho_1}$, where each sequence is generated  from $\prod_{j=1}^n   P_{A|\tilde{U}\tilde{C}C}(\cdot|\tilde{u}_j,\tilde{c}_j,c_j)$.

\item For each $(\tmb{v},\tmb{c},\mathbf{c}) \in \mc{V}^n \times \mc{C}^{n} \times \mc{C}^{n}$ generate independently   $2^{n\rho_2}$ sequences $\mathbf{B}_{[l,i,\tmb{v},\tmb{c},\mathbf{c}]}$,   for $i=1,2,\ldots,2^{n \rho_2}$, where each sequence is generated
  from $\prod_{j=1}^n   P_{B|\tilde{V}\tilde{C}C}(\cdot|\tilde{v}_j,\tilde{c}_j,c_j)$.

\item For each   $(\mathbf{a},\mathbf{b},\mathbf{c},\mathbf{u},\mathbf{v})\in  \mc{A}^n \times \mc{B}^n$ $\times \mc{C}^n \times \mc{U}^n \times \mc{V}^n$ generate  one sequence $\mathbf{X}_{[l,\mathbf{a},\mathbf{b},\mathbf{c},\mathbf{u},\mathbf{v}]}$
  using   $\prod_{i=1}^n   P_{X|ABCUV}(\cdot|a_i,b_i,c_i,u_i,v_i)$.

\item Generate independently sequences $\mathbf{U}[0],\mathbf{V}[0], \mathbf{C}[0], \mathbf{K}[0], \mathbf{X}[0], \mathbf{Y}[0], \mathbf{Z}[0]$ from the product distribution $P_{U,V,C, K, X,Y, Z}^n$.
\end{itemize}

 These sequences are known to all terminals before transmission begins.

\subsection{Encoding Operation} \label{subsec:encoding}
Let $W_0[l]$ denote the common message, and  $W_1[l],W_2[l]$, the private messages for block $l$. These are independent random variables distributed uniformly over $\{0,1,\ldots,2^{nR_0}-1\}$, $\{1,2,\ldots,2^{nR_1}\}$, and $\{1,2,\ldots,2^{nR_2}\}$,
respectively. We set $W_0[0]=W_1[0]=W_2[0]=W_0[L]=W_1[L]=W_2[L]=1$.

For each block $l$, the encoder chooses a quintuple  of sequences $(\mathbf{A}[l],\mathbf{B}[l],\mathbf{C}[l],\mathbf{U}[l],\mathbf{V}[l])$ from
the five codebooks generated above, according to the encoding rule described below. The channel input, and channel output sequences in block $l$ are denoted
$\mathbf{X}[l]$, $\mathbf{Y}[l]$ and $\mathbf{Z}[l]$,
respectively.

 \textbf{Blocks} $l=1,2,3,\ldots,L$: The encoder performs the following sequence of operations.
\begin{itemize}
\item Step $1$: The encoder determines a triplet of indices $G_A[l] \in \{1, \ldots, 2^{n\rho_1} \}$,
$G_B[l] \in \{1, \ldots, 2^{n\rho_2} \}$, and
$G_C[l]) \in \{1, \ldots, 2^{n\rho_0} \} $ such that
\begin{enumerate}
\item $G_C[l] \mod 2^{nR_0} = W_0[l]$, \footnote{This condition corresponds to the role of $C$ in carrying the message $W_0[l]$ common to both receivers.} and
\item The tuple $(\mathbf{U}[l-1],\mathbf{V}[l-1],\mathbf{K}[l-1],\mathbf{C}[l-1])$
is jointly $\e_l$-typical with the triplet of sequences
\ben
\begin{split}
\big\{ \ &\mathbf{C}_{[l,G_C[l],\mathbf{C}[l-1]]}, \\
& \mathbf{A}_{[l,G_A[l],\mathbf{U}[l-1],\mathbf{C}[l-1],\: \mathbf{C}_{[l,G_C[l],\mathbf{C}[l-1]]}]}, \\
& \mathbf{B}_{[l,G_B[l],\mathbf{V}[l-1],\mathbf{C}[l-1],\: \mathbf{C}_{[l,G_C[l],\mathbf{C}[l-1]]}]} \ \big\}
\end{split}
\een
with respect to $P_{\tilde{U},\tilde{V},\tilde{K},\tilde{C},C,A,B}$.\footnote{ If there is more than one triplet satisfying the conditions, the encoder chooses one of them at random.}
\end{enumerate}
If no such index triplet is found, it declares error and  sets
$(G_A[l],G_B[l],G_C[l])=(1,1,1)$.

The encoder then sets
\ben
\begin{split}
& \mathbf{C}[l]=\mathbf{C}_{[l,G_C[l],\mathbf{C}[l-1]]}, \\
& \mathbf{A}[l]= \mathbf{A}_{[l,G_A[l],\mathbf{U}[l-1],\mathbf{C}[l-1],\: \mathbf{C}_{[l,G_C[l],\mathbf{C}[l-1]]}]}, \\
& \mathbf{B}[l]=\mathbf{B}_{[l,G_B[l],\mathbf{V}[l-1],\mathbf{C}[l-1],\: \mathbf{C}_{[l,G_C[l],\mathbf{C}[l-1]]}]}.
\end{split}
\een

\item Step $2$: The encoder chooses a pair of indices $(G_U[l],G_V[l])$ such   that the triplet of sequences
\[
(\mathbf{U}_{[l,G_U[l],\mathbf{C}[l], \mathbf{A}[l]]}, \ \mathbf{V}_{[l,G_V[l],\mathbf{C}[l], \mathbf{B}[l]]},\ \mathbf{A}[l], \ \mathbf{B}[l], \ \mathbf{C}[l] )
\]
 is $\e$-typical with respect to $P_{UVABC}$,   and $\mathbf{U}_{[l,G_U[l],\mathbf{C}[l], \mathbf{A}[l] ]}$ belongs to the $U$-bin with index
  $W_1[l]$, and $\mathbf{V}_{[l,G_V[l],\mathbf{C}[l], \mathbf{B}[l]]}$ belongs to the $V$-bin with index
  $W_2[l]$. If no such index pair is found, it declares error and sets $(G_U[l],G_V[l])=(1,1)$.

The encoder then sets
$\mathbf{U}[l]=\mathbf{U}_{[l,G_U[l],\mathbf{C}[l], \mathbf{A}[l]]}$, $\mathbf{V}[l]=\mathbf{V}_{[l,G_V[l],\mathbf{C}[l], \mathbf{B}[l]]}$, and
$\mathbf{X}[l]=\mathbf{X}_{[l,\mathbf{A}[l],\mathbf{B}[l],\mathbf{C}[l],\mathbf{U}[l],\mathbf{V}[l]]}$. It transmits $\mathbf{X}[l]$ as the channel input sequence for block $l$.

\item Step $3$: The broadcast channel produces $(\mathbf{Y}[l],\mathbf{Z}[l])$.

\item Step $4$: After receiving $(\mathbf{S}[l])$ via the feedback link, the encoder
  sets  $\mathbf{K}[l]=(\mathbf{A}[l],\mathbf{B}[l],\mathbf{S}[l])$.
\end{itemize}
\subsection{Decoding Operation}
\noindent \textbf{Block $1$}: The objective at the end of this block is to decode the common message $W_0[1]$ at both receivers.
\begin{itemize}
\item The first decoder receives $\mathbf{Y}[1]$, and the second decoder receives $\mathbf{Z}[1]$.

\item The first decoder determines the unique index pair   $(\hat{G}_{C1}[1],\hat{G}_A[1])$ such that the tuples
$(\mathbf{C}[0],\mathbf{A}[0],\mathbf{U}[0], \mathbf{Y}[0])$  and  $(\bar{\mathbf{C}}_1[1], \mathbf{A}_{[1,\hat{G}_A[1],\mathbf{U}[0], \mathbf{C}[0],\bar{\mathbf{C}}_1[1]]}, \mathbf{Y}[1])$ are jointly $\e_l$-typical with respect to $P_{\tilde{C} \tilde{A}\tilde{U}\tilde{Y} CAY}$,
where $\bar{\mathbf{C}}_1[1] \triangleq \mathbf{C}_{[1,\hat{G}_{C1}[1],\mathbf{C}[0]]}$. Note that $\bar{\mathbf{C}}_1[1]$ is the estimate of $\mathbf{C}[1]$ at the first decoder.

If not successful in this operation, the first decoder declares an error and  sets $(\hat{G}_{C1}[1],\hat{G}_A[1])=(1,1)$, and
$\bar{\mathbf{C}}_1[1] \triangleq  \mathbf{C}_{[1,\hat{G}_{C1}[1],\mathbf{C}[0]]}$.

\item The first decoder outputs $\hat{W}_0[1]=\hat{G}_{C1}[1]\mod 2^{nR_0}$, and sets
\[
\bar{\mathbf{A}}[1]=\mathbf{A}_{[1,\hat{G}_A[1],\mathbf{U}[0],\mathbf{C}[0], \bar{\mathbf{C}}_1[1]]}.
\]
$\bar{\mathbf{A}}[1]$ is the first decoder's estimate of $\mathbf{A}[1]$.

\item The second decoder determines the unique index pair   $(\hat{G}_{C2}[1],\hat{G}_B[1])$ such that the tuples $(\mathbf{C}[0],\mathbf{B}[0], \mathbf{V}[0], \mathbf{Z}[0])$  and   $(\bar{\mathbf{C}}_2[1], \mathbf{B}_{[1,\hat{G}_B[1],\mathbf{V}[0],\mathbf{C}[0],\bar{\mathbf{C}}_2[1]]}, \mathbf{Z}[1])$
are jointly $\e_l$-typical with respect to $P_{\tilde{C}\tilde{B}\tilde{V} \tilde{Z}CBZ}$, where $\bar{\mathbf{C}}_2[1] \triangleq  \mathbf{C}_{[1,\hat{G}_{C2}[1],\mathbf{C}[0]]}$. Note that $\bar{\mathbf{C}}_2[1]$ is the estimate of $\mathbf{C}[1]$, at the second decoder.

If not successful in this operation, the second decoder declares an error and  sets $(\hat{G}_{C2}[1],\hat{G}_B[1])=(1,1)$, and
$\bar{\mathbf{C}}_2[1] \triangleq  \mathbf{C}_{[1,\hat{G}_{C2}[1],\mathbf{C}[0]]}$.

\item The second decoder outputs $\bar{W}_0[1]=\hat{G}_{C2}[1]\mod 2^{nR_0}$, and sets
\[
\bar{\mathbf{B}}[1]=\mathbf{B}_{[1,\hat{G}_B[1],\mathbf{V}[0],\mathbf{C}[0], \bar{\mathbf{C}}_2[1]]}.
\]
$\bar{\mathbf{B}}[1]$ is the second decoder's estimate of $\mathbf{B}[1]$.
\end{itemize}

\noindent \textbf{Block} $l, l=2,3,\ldots,L$: The objective at the end of block $l$ is for receiver $1$ to decode $(W_0[l], W_1[l-1])$ and
for receiver $2$ to decode $(W_0[l], W_2[l-1])$.
\begin{itemize}
\item The first decoder receives $\mathbf{Y}[l]$ and the second decoder receives $\mathbf{Z}[l]$.
\item The first decoder determines the unique index triplet $(\hat{G}_{C1}[l],\hat{G}_A[l],\hat{G}_U[l-1])$ such that the tuples
\ben
\begin{split}
& (\bar{\mathbf{C}}_1[l-1],\bar{\mathbf{A}}[l-1], \bar{\mathbf{U}}[l-1], \mathbf{Y}[l-1]) \ \text{ and } \\
& (\bar{\mathbf{C}}_1[l], \mathbf{A}_{[l,\hat{G}_A[l],\bar{\mathbf{U}}[l-1], \bar{\mathbf{C}}_1[l-1],\bar{\mathbf{C}}_1[l]]}, \mathbf{Y}[l])
\end{split}
\een
are jointly $\e_l$-typical with respect to $P_{\tilde{C} \tilde{A}\tilde{U}\tilde{Y} CAY}$, where
\ben
\begin{split}
& \bar{\mathbf{U}}[l-1] \triangleq \mathbf{U}_{[(l-1),\hat{G}_U[l-1],\bar{\mathbf{C}}_1[l-1], \bar{\mathbf{A}}[l-1]]}, \\
& \bar{\mathbf{C}}_1[l] \triangleq  \mathbf{C}_{[l,\hat{G}_{C1}[l],\bar{\mathbf{C}}_1[l-1]]}.
\end{split}
\een

If not successful in this operation, the first decoder  declares an error and sets $(\hat{G}_{C1}[l],\hat{G}_A[l],\hat{G}_U[l-1])=(1,1,1)$, and $\bar{\mathbf{U}}[l-1] = \mathbf{U}_{[(l-1),1,\bar{\mathbf{C}}_1[l-1], \bar{\mathbf{A}}[l-1]]}, \
\bar{\mathbf{C}}_1[l] =  \mathbf{C}_{[l,1,\bar{\mathbf{C}}_1[l-1]]}$. Note that $\bar{\mathbf{U}}[l-1]$ and $\bar{\mathbf{C}}_1[l]$ are
the estimates of $\mathbf{U}[l-1]$ and $\mathbf{C}[l]$, respectively, at the first decoder.

\item The first decoder  then outputs $\hat{W}_0[l]=\hat{G}_{C1}[l]\mod 2^{nR_0}$, and
$\hat{W}_1[l-1]$ as the index of $U$-bin that contains the
sequence $\mathbf{U}_{[(l-1),\hat{G}_U[l-1],\bar{\mathbf{C}}_1[l-1], \bar{\mathbf{A}}[l-1]]}$. The decoder  sets
\[
\bar{\mathbf{A}}[l]=\mathbf{A}_{[l,\hat{G}_A[l],\bar{\mathbf{U}}[l-1],\bar{\mathbf{C}}_1[l-1],
\bar{\mathbf{C}}_1[l]]}.
\]
$\bar{\mathbf{A}}[l]$ is the first decoder's estimate of $\mathbf{A}[l]$.

\item The second decoder determines the unique index triplet  $(\hat{G}_{C2}[l],\hat{G}_B[l],\hat{G}_V[l-1])$ such that the tuples
\ben
\begin{split}
& (\bar{\mathbf{C}}_2[l-1],\bar{\mathbf{B}}[l-1], \bar{\mathbf{V}}[l-1], \mathbf{Z}[l-1]) \ \text{ and } \\
& (\bar{\mathbf{C}}_2[l], \mathbf{B}_{[l,\hat{G}_B[l],\bar{\mathbf{V}}[l-1], \bar{\mathbf{C}}_2[l-1],\bar{\mathbf{C}}_2[l]]}, \mathbf{Z}[l])
\end{split}
\een
are jointly $\e_l$-typical with respect to $P_{\tilde{C} \tilde{B}\tilde{V}\tilde{Z} CBZ}$, where
\ben
\begin{split}
& \bar{\mathbf{V}}[l-1] \triangleq \mathbf{V}_{[(l-1),\hat{G}_V[l-1],\bar{\mathbf{C}}_2[l-1], \bar{\mathbf{B}}[l-1]]}, \\
& \bar{\mathbf{C}}_2[l] \triangleq  \mathbf{C}_{[l,\hat{G}_{C2}[l],\bar{\mathbf{C}}_2[l-1]]}.
\end{split}
\een

If not successful in this operation, the second decoder declares an error and sets
$(\hat{G}_{C2}[l],\hat{G}_B[l],\hat{G}_V[l-1])=(1,1,1)$, and
$\bar{\mathbf{V}}[l-1] = \mathbf{V}_{[(l-1),1,\bar{\mathbf{C}}_2[l-1], \bar{\mathbf{B}}[l-1]]}, \
\bar{\mathbf{C}}_2[l] =  \mathbf{C}_{[l,1,\bar{\mathbf{C}}_2[l-1]]}$.
Note that $\bar{\mathbf{V}}[l-1]$ and $\bar{\mathbf{C}}_2[l]$ are the estimates of $\mathbf{V}[l-1]$ and $\mathbf{C}[l]$, respectively,
at the second decoder.

\item The second decoder then outputs $\bar{W}_0[l]=\hat{G}_{C2}[l]\mod 2^{nR_0}$,
and $\hat{W}_2[l-1]$ as the index of $V$-bin that contains the
sequence $\mathbf{V}_{[(l-1),\hat{G}_V[l-1],\bar{\mathbf{C}}_2[l-1], \bar{\mathbf{B}}[l-1]]}$. The decoder sets
\[
\bar{\mathbf{B}}[l]=\mathbf{B}_{[l,\hat{G}_B[l],\bar{\mathbf{V}}[l-1],\bar{\mathbf{C}}_2[l-1],
\bar{\mathbf{C}}_2[l]]}.
\]
$\bar{\mathbf{B}}[l]$ is the second decoder's estimate of $\mathbf{B}[l]$.
\end{itemize}

\subsection{Error Analysis}
Let $\mc{E}[0]$ denote the event that
$(\mathbf{U}[0],\mathbf{K}[0],\mathbf{V}[0],\mathbf{C}[0])$ is not $\e[0]$-typical
with respect to $P_{UKVC}$. By Property 0, we have $\text{Pr}[\mc{E}[0]] \leq \e$
for all sufficiently large $n$.

 \textbf{Block $1$:}
The error event in Block $1$ can be expressed as $\mc{E}[1]=\mc{E}_1[1] \cup \mc{E}_2[1] \cup
\mc{E}_3[1] \cup \mc{E}_4[1] \cup \mc{E}_5[1]$ where
\begin{itemize}
\item[-] $\mc{E}_1[1]$ is the event that the encoder declares error  in
step $1$ of encoding (described in Section \ref{subsec:encoding}),
\item[-] $\mc{E}_2[1]$ is the event that the encoder declares error  in
step $2$ of encoding,
\item[-] $\mc{E}_3[1]$ is the event that the tuples
$(\mathbf{U}[0],\mathbf{V}[0],\mathbf{K}[0],\mathbf{C}[0])$   {and} $(\mathbf{U}[1],\mathbf{V}[1], \mathbf{K}[1], \mathbf{C}[1])$
are not jointly $\e_1$-typical with respect to $P_{\tilde{U}\tilde{V}\tilde{K}\tilde{C}UVKC}$,
\item[-] $\mc{E}_4[1]$ is the event that $(\hat{G}_{C1}[1],\hat{G}_A[1]) \neq (G_C[1],G_A[1])$, and
$\mc{E}_5[1]$ is the event that $(\hat{G}_{C2}[1],\hat{G}_B[1]) \neq (G_C[1],G_B[1])$.
\end{itemize}

\begin{lem}[Covering lemma]
\label{lem:error1}
$\text{Pr}[\mc{E}_1[1] \mid \mc{E}[0]^c] \leq \e$ for all sufficiently large $n$ if
$R_0$, $\rho_0,\rho_1$, and $\rho_2$ satisfy
\begin{align}
&\rho_0  > I(\tilde{U} \tilde{V} \tilde{K};C|\tilde{C})+R_0 + \delta(\e_1) \label{eq:rate21} \\
& \rho_0+\rho_1 > I(\tilde{V}\tilde{K} ; A|C \tilde{C} \tilde{U}) +I(\tilde{U} \tilde{V} \tilde{K} ;C|\tilde{C}) +R_0 + \delta(\e_1)\\
& \rho_0+\rho_2 > I(\tilde{U} \tilde{K} ;B|C \tilde{C} \tilde{V}) +I(\tilde{U} \tilde{V} \tilde{K};C|\tilde{C}) +R_0 + \delta(\e_1)\\
& \rho_0+\rho_1+\rho_2  >  I(\tilde{V}\tilde{K} ; A|C \tilde{C} \tilde{U}) + I(\tilde{U} \tilde{K} ;B|C \tilde{C} \tilde{V}) \nonumber \\
& \ \ +I(A;B| \tilde{U} \tilde{V} \tilde{K} C \tilde{C})+I(\tilde{U} \tilde{V} \tilde{K};C|\tilde{C})+R_0 + \delta(\e_1) \label{eq:rate31}
\end{align}
\end{lem}
\begin{IEEEproof} The proof of this covering lemma is the same as that of \cite[Lemma 14.1]{ElGKimBook}, with $(\tilde{U}, \tilde{K})$ and $(\tilde{V}, \tilde{K})$ assuming the roles of the two sources being covered. \end{IEEEproof}

\begin{lem}
$\text{Pr}[\mc{E}_2[1] \mid \mc{E}[0]^c] \leq \e$ for all sufficiently large $n$ if
$R'_1,R'_2$, and $R_1,R_2$ satisfy
\begin{equation}
\begin{split}
R'_1+R'_2-R_1-R_2 & > H(U|AC) + H(V|BC)\\
&  \quad - H(UV|ABC) + \delta(\e_1)  \label{eq:rate11}
\end{split}
\end{equation}
\label{lem:error2}
\end{lem}
\begin{IEEEproof} This is very similar to a standard covering lemma used for bounding the probability of encoding error in Marton's coding scheme, a proof of which can be found in \cite{ElGMuelen81, Cover98} or \cite{ElGKimBook}.
\end{IEEEproof}

From Property $1$ of typical sequences, it follows that $\text{Pr}[\mc{E}_3[1] \mid \mc{E}_1[1]^c,\mc{E}_2[1]^c,\mc{E}[0]^c] \leq \e$ for all sufficiently
large $n$.
\begin{lem} \label{lem:error2a}
$\text{Pr}[\mc{E}_4[1] \cup \mc{E}_5[1] | \mc{E}_3[1]^c, \mc{E}_2[1]^c, \mc{E}_1[1]^c, \mc{E}[0]^c]
\leq 2\e$, if
\begin{align}
&\rho_0+\rho_1 < I(C;Y\tilde{A} \tilde{Y}\tilde{U}|\tilde{C})
+I(A;Y\tilde{A}\tilde{Y}| \tilde{U} \tilde{C} C) -3\delta(\e_1) \\
&\rho_0+\rho_2 < I(C;Z \tilde{B} \tilde{Z}\tilde{V}|\tilde{C})
+I(B;Z\tilde{B}\tilde{Z}| \tilde{V} \tilde{C} C) -3\delta(\e_1) \\
&\rho_1 < I(A;Y \tilde{A} \tilde{Y}| \tilde{U} \tilde{C} C) -3\delta(\e_1)\\
&\rho_2 < I(B;Z \tilde{B} \tilde{Z}| \tilde{V} \tilde{C} C) -3\delta(\e_1)
\end{align}
\end{lem}
\begin{IEEEproof} The proof is a special case of that of Lemma \ref{lem:error3} given below.
\end{IEEEproof}

Hence $P[\mc{E}[1] \mid \mc{E}[0]^c] < 5 \e$ if the conditions given in Lemmas \ref{lem:error1}, \ref{lem:error2}, and \ref{lem:error2a} are satisfied.
This implies that $\bar{\mathbf{A}}[1]=\mathbf{A}[1]$, $\bar{\mathbf{C}}_1[1]=\bar{\mathbf{C}}_2[1]=\mathbf{C}[1]$,  and similarly $\bar{\mathbf{B}}[1]=\mathbf{B}[1]$ with high probability.

\textbf{Block} $l$, $l= 2,3,\ldots,L$: The error event in block $l$  can be expressed as $\mc{E}[l]=\cup_{i=1}^5 \mc{E}_i[l]$ where
\begin{itemize}
\item[-] $\mc{E}_1[l]$ is the event that the encoder declares error  in step $1$ of encoding
\item[-] $\mc{E}_2[l]$  is the event that the encoder declares error  in  step $2$ of encoding,
\item[-] $\mc{E}_3[l]$ is the event that the tuples
$(\mathbf{U}[l-1],\mathbf{V}[l-1],\mathbf{K}[l-1],\mathbf{C}[l-1])$ {and} $(\mathbf{U}[l],\mathbf{V}[l],\mathbf{K}[l],\mathbf{C}[l])$
are not jointly $\e_l$-typical with respect to $P_{\tilde{U}\tilde{V}\tilde{K}\tilde{C}UVKC}$,
\item[-] $\mc{E}_4[l]$ is the event that
\[ \{ (\hat{G}_{C1}[l],\hat{G}_A[l],\hat{G}_U[l-1]) \neq (G_C[l],G_A[l],G_U[l-1]) \}\] and  $\mc{E}_5[l]$ is the event that
\[ \{(\hat{G}_{C2}[l],\hat{G}_B[l],\hat{G}_V[l-1]) \neq (G_C[l],G_B[l],G_V[l-1]) \}. \]
\end{itemize}

Using arguments similar to those used for  Block $1$, one can show that if $\rho_0,\rho_1,\rho_2$, $R'_1,R'_2$, $R_1$ and $R_2$ satisfy the conditions given in \eqref{eq:rate11} and \eqref{eq:rate21}-\eqref{eq:rate31} with $\e[l-1]$ replaced with $\e_l$,
then  for all sufficiently large $n$,
\[
\text{Pr}[\mc{E}_1[l] \cup \mc{E}_2[l] \cup \mc{E}_3[l] \mid \cap_{k=0}^{l-1} \mc{E}[k]^c] \leq 3\e.
\]

\begin{lem}[Packing lemma]
\[\text{Pr}[\mc{E}_4[l] \cup \mc{E}_5[l] \mid \mc{E}_3[l]^c, \mc{E}_2[l]^c, \mc{E}_1[l]^c, \cap_{k=0}^{l-1} \mc{E}[k]^c]
\leq 2\e, \text{ if} \]
\begin{align}
& R'_1 + \rho_0 + \rho_1  < I(\tilde{U};Y\tilde{Y} | \tilde{A} \tilde{C}) + I(C;Y \tilde{A} \tilde{Y} \tilde{U}|\tilde{C}) \nonumber \\
& \qquad \qquad \qquad  \quad + I(A;Y\tilde{A}\tilde{Y} |\tilde{U}  \tilde{C} C) - 3\delta(\e_l) \label{eq:rate41} \\
& R'_1+\rho_1 < I(\tilde{U};Y  \tilde{Y} C | \tilde{A} \tilde{C}) +I(A;Y\tilde{A} \tilde{Y} |\tilde{U} \tilde{C} C) - 3\delta(\e_l)\\
& R'_2 + \rho_0 + \rho_2 < I(\tilde{V};Z\tilde{Z} | \tilde{B} \tilde{C}) +I(C;Z \tilde{B} \tilde{Z} \tilde{V}|\tilde{C})  \nonumber \\
& \qquad \qquad \qquad  \quad + I(B;Z\tilde{B} \tilde{Z} |\tilde{V} \tilde{C} C) - 3\delta(\e_l) \\
& R'_2+\rho_2 < I(\tilde{V};Z \tilde{Z} C| \tilde{B}  \tilde{C}) +I(B;Z\tilde{B} \tilde{Z} |\tilde{V} \tilde{C} C) - 3\delta(\e_l) \\
& \rho_0+\rho_1 < I(C;Y\tilde{A} \tilde{Y}\tilde{U}|\tilde{C}) +I(A;Y\tilde{A}\tilde{Y}| \tilde{U} \tilde{C} C) - 3\delta(\e_l) \\
& \rho_0+\rho_2 < I(C;Z \tilde{B} \tilde{Z}\tilde{V}|\tilde{C}) +I(B;Z\tilde{B}\tilde{Z}| \tilde{V} \tilde{C} C)- 3\delta(\e_l)  \\
& \rho_1 < I(A;Y \tilde{A} \tilde{Y}| \tilde{U} \tilde{C} C) - 3\delta(\e_l)\\
& \rho_2 < I(B;Z \tilde{B} \tilde{Z}| \tilde{V} \tilde{C} C) - 3\delta(\e_l) \label{eq:rate51}
\end{align}
\label{lem:error3}
\end{lem}
\begin{IEEEproof} See Appendix \ref{app:lem_error3}. \end{IEEEproof}

Hence $\text{Pr}[\mc{E}[l] \mid \cap_{k=0}^{l-1} \mc{E}[k]^c] < 5 \e$. Under the event $\cap_{k=0}^{l} \mc{E}[k]^c$, we have
$\bar{\mathbf{A}}[l]=\mathbf{A}[l]$, $\bar{\mathbf{C}}_1[l]=\bar{\mathbf{C}}_2[l]=\mathbf{C}[l]$,  and $\bar{\mathbf{B}}[l]=\mathbf{B}[l]$.

\emph{Overall Probability of Decoding Error}:
The analysis above shows that the probability of decoding error over $L$ blocks satisfies
\[
\text{Pr}[\mc{E}] = Pr \left[ \cup_{l=0}^{L} \mc{E}[l] \right] \leq 5 \e L
\]
if the  conditions given in \eqref{eq:rate11}, \eqref{eq:rate21}-\eqref{eq:rate31} and \eqref{eq:rate41}-\eqref{eq:rate51} are satisfied with $\delta(\e_1)$
and $\delta(\e_l)$ are  replaced with $\theta$, where $\theta=\sum_{l=1}^{L} \delta(\e_l)$. This implies that the rate region given by \eqref{eq:rate1}, \eqref{eq:rate2}-\eqref{eq:rate3}, \eqref{eq:rate4}-\eqref{eq:rate5} is achievable. By applying Fourier-Motzkin elimination to these equations, we obtain that the rate region given in the statement of the theorem is achievable. The details of this elimination are omitted since they are elementary, but somewhat tedious.

\section{Conclusion} \label{sec:conc}
We have derived a single-letter rate region for the two-user broadcast channel with feedback. Using the Marton coding scheme as the starting
point, our scheme has a block-Markov  structure and uses three additional random variables ($A,B,C$) to cover the correlated information generated at the end of each block.

The proposed region was used to compute achievable rates for the AWGN channel with noisy feedback. In particular, it was shown that sum rates higher than the no-feedback sum capacity could be achieved even with noisy feedback to only one receiver. In all the examples including the AWGN channel, the improvement over the no-feedback region was obtained using a simplified version of the rate region with the resolution information carried only by the common random variable $C$. An open question is whether the AWGN sum-rate with noisy feedback can be improved by sending resolution information via $A$ and $B$ as well. Since the resolution information used for the two receivers are correlated Gaussians, the results of \cite{TianDS11,BrossLT10} suggest that this may be possible.

The key to obtaining a single-letter characterization was to impose a constraint on the Markov kernel connecting the distribution of the random variables across successive blocks. A similar idea was used in \cite{RamjiMAC11} for multiple-access channels with feedback. This approach to harnessing correlated information is quite general, and it is likely that it can be used to obtain improved rate regions for other  multi-user channels with feedback such as interference and relay channels.

\appendices

\section{Mutual Information terms  for the AWGN example}
\label{app:awgn_compute}
With the joint distribution described in Section \ref{subsec:awgn}, we first compute the following quantities.
\begin{align}
M_u  \triangleq E[\Delta^2] & = \alpha P_1 \left(\frac{\sigma^2 + \bar{\beta}\bar{\alpha}P_1}{P_1+\sigma^2}\right)^2
+ \bar{\alpha}P_1 \left(\frac{\beta \sigma^2 - \alpha \bar{\beta}P_1 }{P_1+\sigma^2}\right)^2 \nonumber \\
& +  \left(\frac{\alpha P_1 + \beta \bar{\alpha} P_1}{P_1+\sigma^2}\right)^2 \frac{\sigma^4}{\sigma^2 + \sigma_f^2},\\
E[\Delta \tilde{V}] &=   \frac{\bar{\alpha} P_1 (\beta \sigma^2 - \alpha \bar{\beta}P_1)}{P_1+\sigma^2}, \\
E[\Delta \tilde{U}] & = \frac{\alpha P_1 (\sigma^2 + \bar{\alpha}\bar{\beta} P_1)}{P_1+\sigma^2} +
\frac{\beta \bar{\alpha} P_1 (\beta \sigma^2 - \alpha \bar{\beta}P_1)}{P_1+\sigma^2}, \\
E[\Delta \tilde{Z}] & = \frac{P_1 \sigma^2 (\alpha + \bar{\alpha}\beta)}{P_1+\sigma^2}\left( 1 - \frac{\sigma^2 \rho}{\sigma^2 + \sigma_f^2} \right),   \\
E[\Delta \tilde{Y}] & =  \frac{P_1 \sigma^2 (\alpha + \bar{\alpha}\beta)}{P_1+\sigma^2}\left( \frac{\sigma_f^2}{\sigma^2 + \sigma_f^2} \right).
\end{align}
We next compute the conditional variances in terms of which the mutual information quantities are expressed.
\begin{align}
& \text{var}(\tilde{T}_1|\tilde{C}_0\tilde{V}) = 1 - \frac{(E[\Delta \tilde{V}])^2}{M_u \bar{\alpha}P_1} \\
& \text{var}(\tilde{T}_1|\tilde{C}_0\tilde{U}) = 1 - \frac{(E[\Delta \tilde{U}])^2}{M_u (\alpha P_1 + \beta^2 \bar{\alpha} P_1)} \\
& \text{var}(\tilde{T}_1|\tilde{C}_0\tilde{Z}) = 1 - \frac{(E[\Delta \tilde{Z}])^2}{M_u (P_1+\sigma^2)} \\
& \text{var}(\tilde{T}_1|\tilde{C}_0\tilde{Y}) = 1 - \frac{(E[\Delta \tilde{Y}])^2}{M_u (P_1+\sigma^2)} \\
& \text{var}(\tilde{T}_1|\tilde{C}_0 \tilde{V} \tilde{Z})  = 1 - \frac{1}{M_u}\left(\frac{a_1 E[\Delta \tilde{V}] + b_1 E[\Delta \tilde{Z}]}{\bar{\alpha}P_1(\alpha P_1 + \sigma^2)} \right)\\
& \text{var}(\tilde{T}_1|\tilde{C}_0 \tilde{U} \tilde{Y}) = \nonumber \\
& 1 - \frac{1}{M_u}\left(\frac{a_2 E[\Delta \tilde{U}] + b_2 E[\Delta \tilde{Y}]}
{(P_1+\sigma^2)({\alpha}P_1 + \beta^2 \bar{\alpha} P_1) - ({\alpha}P_1 + \beta \bar{\alpha} P_1)^2 } \right)
\end{align}
where
\be
\begin{split}
& a_1 = E[\Delta \tilde{V}] (P_1 + \sigma^2)  - E[\Delta \tilde{Z}] \bar{\alpha}P_1, \\
& b_1=E[\Delta \tilde{Z}] \bar{\alpha}P_1  - E[\Delta \tilde{V}] \bar{\alpha}P_1, \\
& a_2 = E[\Delta \tilde{U}] (P_1 + \sigma^2)  - E[\Delta \tilde{Y}] ({\alpha}P_1 + \beta \bar{\alpha} P_1), \\
&   b_2= E[\Delta \tilde{Y}] ({\alpha}P_1 + \beta^2 \bar{\alpha} P_1) - E[\Delta \tilde{U}] ({\alpha}P_1 + \beta \bar{\alpha} P_1).
\end{split}
\ee
Finally, the mutual information terms are calculated to be

\begin{align*}
& I(U;V)= \frac{1}{2} \log\left(1+ \frac{\beta^2 \bar{\alpha}}{\alpha}\right),  \\
& I(U;Y|C_0) =  \frac{1}{2} \log\left(\frac{(P_1+\sigma^2)(\alpha + \beta^2 \bar{\alpha})}{(P_1+\sigma^2)(\alpha+\beta^2\bar{\alpha}) - P_1(\alpha + \beta \bar{\alpha})^2} \right)\\
& I(V;Z|C_0) = \frac{1}{2} \log\left( \frac{P_1+\sigma^2}{\alpha P_1 + \sigma^2}\right),
\\
& I(C_0;\tilde{U}\tilde{S}|\tilde{C}_0 \tilde{V}) = \frac{1}{2} \log \left( 1 + \frac{(1-D)}{D} \text{var}(\tilde{T}_1|\tilde{C}_0 \tilde{V}) \right), \\ & I(C_0;\tilde{V}\tilde{S}|\tilde{C}_0 \tilde{U}) = \frac{1}{2} \log \left( 1 + \frac{(1-D)}{D} \text{var}(\tilde{T}_1|\tilde{C}_0 \tilde{U}) \right),    \\
& I(C_0;\tilde{Y}|\tilde{C}_0 \tilde{U}) =  \frac{1}{2} \log \left(\frac{(1-D) \text{var}(\tilde{T}_1|\tilde{C}_0 \tilde{U}) + D}{(1-D) \text{var}(\tilde{T}_1|\tilde{C}_0 \tilde{U} \tilde{Y}) + D} \right),  \\
& I(C_0;\tilde{Z}|\tilde{C}_0 \tilde{V}) =  \frac{1}{2} \log \left( \frac{(1-D) \text{var}(\tilde{T}_1|\tilde{C}_0 \tilde{V}) + D}{(1-D) \text{var}(\tilde{T}_1|\tilde{C}_0 \tilde{V} \tilde{Z}) + D}\right),
\\
& I(C_0;\tilde{U}|\tilde{C}_0 \tilde{Y}) =  \frac{1}{2} \log \left(\frac{(1-D) \text{var}(\tilde{T}_1|\tilde{C}_0 \tilde{Y}) + D}{(1-D) \text{var}(\tilde{T}_1|\tilde{C}_0 \tilde{U} \tilde{Y}) + D} \right),  \\
& I(C_0;\tilde{V}|\tilde{C}_0 \tilde{Z}) =  \frac{1}{2} \log \left( \frac{(1-D) \text{var}(\tilde{T}_1|\tilde{C}_0 \tilde{Z}) + D}{(1-D) \text{var}(\tilde{T}_1|\tilde{C}_0 \tilde{V} \tilde{Z}) + D}\right), \\
& I(C_0;Y|\tilde{C}_0 \tilde{Y})= \\
&\frac{1}{2} \log \left(1+  \frac{(P-P_1)( (1-D)\text{var}(\tilde{T}_1|\tilde{C}_0 \tilde{Y}) + D )}{P_1+\sigma^2} \right),\\
&I(C_0;Z|\tilde{C}_0 \tilde{Z})= \\
& \frac{1}{2} \log \left(1+  \frac{(P-P_1)( (1-D)\text{var}(\tilde{T}_1|\tilde{C}_0 \tilde{Z}) + D )}{P_1+\sigma^2} \right),\\
&I(C_0;Y|\tilde{C}_0 \tilde{U} \tilde{Y})= \\
& \frac{1}{2} \log \left(1+  \frac{(P-P_1)( (1-D)\text{var}(\tilde{T}_1|\tilde{C}_0 \tilde{U} \tilde{Y}) + D )}{P_1+\sigma^2} \right), \\
& I(C_0;Z|\tilde{C}_0 \tilde{V} \tilde{Z})= \\
&\frac{1}{2}\log \left(1 + \frac{(P-P_1)((1-D)\text{var}(\tilde{T}_1|\tilde{C}_0 \tilde{V} \tilde{Z}) +D)}{P_1+\sigma^2} \right).
\end{align*}
\section{Comparison with Shayevitz-Wigger Region} \label{app:comparison}
In Theorem \ref{thm:mainthm_a}, set $A=\tilde{V}_1, B=\tilde{V}_2, C=(\tilde{V}_0, W)$ and consider joint distributions over two blocks of the form
\be P_{\tilde{U} \tilde{V} \tilde{W} \tilde{X} \tilde{Y} \tilde{Z} \tilde{S}} \ Q_{\tilde{V}_0 \tilde{V}_1 \tilde{V}_2|\tilde{U} \tilde{V} \tilde{W} \tilde{S}} \ P_{WUV} \ P_{X|WUV} \ P_{YZS|X}.  \label{eq:sw_joint} \ee
If we set $Q_{\tilde{V}_0 \tilde{V}_1 \tilde{V}_2|\tilde{U} \tilde{V} \tilde{W} \tilde{S}} = P_{V_0 | UVWS}  P_{V_1 | V_0 U V W S}  P_{V_2 |V_0 U V W S} $, the joint distribution is identical to that of the Shayevitz-Wigger region. With this distribution, Theorem \ref{thm:mainthm_a} yields the following. (The parts in bold indicate the corresponding constraints of the Shayevitz-Wigger region.)
\begin{align}
& R_0  < \{ \mc{T}_1, \mc{T}_2\} \label{eq:sw_eval1}\\
& R_0 + R_1 <  \mathbf{I(UW; YV_1) - I(UVWS ; V_0V_1| Y)} \nonumber \\
& \qquad \qquad  + I(V_0; UW|V_1Y) \\
& R_0 + R_2 < \mathbf{I(VW; ZV_2) - I(UVWS ; V_0V_2 | Y)} \nonumber \\
& \qquad \qquad + I(V_0; VW|V_2Z) \\
& R_0 + R_1 + R_2 < \nonumber \\
&\  \mathbf{I(UW; YV_1) + I(V;ZV_2|W) - I(U;V|W)} \nonumber \\
&\  \mathbf{- I(UVWS; V_0V_1|Y) - I(UVWS;V_2|V_0Z)}  \\
&\  + I(V_0; UW|V_1 Y) + I(V_0; V | V_2 WZ) + I(V_2; W|ZV_0) \nonumber\\
&  R_0 + R_1 + R_2 < \nonumber \\
&\ \mathbf{I(VW; ZV_2) + I(U;YV_1|W) - I(U;V|W)} \nonumber\\
&\ \mathbf{ - I(UVWS; V_0V_2|Z) - I(UVWS;V_1|V_0Y)} \\
&\ + I(V_0; VW|V_2 Z) + I(V_0; U | V_1 WY) + I(V_1; W|YV_0)\nonumber\\
& 2R_0 + R_1 + R_2 < \nonumber \\
&\ \mathbf{I(UW; YV_1) + I(VW; ZV_2) - I(U;V|W)}\nonumber \\
&\ \mathbf{- I(UVWS ; V_0V_1 | Y) - I( UVWS ; V_0V_2 | Z)} \label{eq:sw_eval6}\\
&\ + I(V_0; UW|V_1Y)  + I(V_0; VW|V_2Z)\nonumber
\end{align}
where
\ben
\begin{split}
\mc{T}_1 & =  I(W;Y) + I(V_1V_0 ; Y | WU) - I(VS; V_1V_0|WU)\\
& \quad + I(V_2;Z|WV V_0) - I(US; V_2|W V V_0), \\
\mc{T}_2 & = I(W;Z) + I(V_2V_0 ; Z | WV) - I(US; V_2V_0|WV) \\
& \quad + I(V_1;Y|WU V_0) - I(V S; V_1|W U V_0). \\
\end{split}
\een
\section{Proof of Lemma \ref{lem:error3}} \label{app:lem_error3}
We show through induction that if $\text{Pr}[\mc{E}_4[k]] < \epsilon$ for $k=1, \ldots, l-1$, then $\text{Pr}(\mc{E}_4[l]) < \epsilon$ if the conditions in the statement of the lemma are satisfied.

For conciseness, let $\mc{F}$ denote the event $( \cap_{k=0}^{l-1} \mc{E}[k]^c \cap \mc{E}_1[l]^c \cap \mc{E}_2[l]^c \cap \mc{E}_3[l]^c)$. Note that $\mc{F}$ is the conditioning event in the statement of Lemma \ref{lem:error3}; hence  $\bar{\mathbf{C}}_1[l-1]=\bar{\mathbf{C}}_2[l-1]=\mathbf{C}[l-1]$, $\bar{\mathbf{A}}[l-1]=\mathbf{A}[l-1]$ and
$\bar{\mathbf{B}}[l-1]=\mathbf{B}[l-1]$.
Recall that given $\mathbf{C}[l-1], \mathbf{A}[l-1]$ and the indices $G_C[l], G_A[l], G_U[l-1]$,  the following sequences are determined:
\begin{equation}
\begin{split}
& \mathbf{U}[l-1]  = \mathbf{U}_{[l-1,G_U[l-1],{\mathbf{C}}[l-1], {\mathbf{A}}[l-1]]}, \\
\mathbf{C}[l]  =   & \mathbf{C}_{[l,G_C[l],{\mathbf{C}}[l-1]]}, \quad
\mathbf{A}[l]  = \mathbf{A}_{[l,G_A[l],\mathbf{U}[l-1],\mathbf{C}[l-1], \mathbf{C}[l]]}.
\end{split}
\end{equation}
Define the  following indicator random variable:
$\psi(i,j,k)=1$ if the tuples
\ben
\begin{split}
& (\mathbf{U}_{[l-1,k,{\mathbf{C}}[l-1], \ \mathbf{A}[l-1]]},\ {\mathbf{A}}[l-1], \ \mathbf{Y}[l-1], \ \mathbf{C}[l-1])
\ \mbox{ and } \\
& (\mathbf{C}_{[l,i,{\mathbf{C}}[l-1]]},\mathbf{Y}[l],
\mathbf{A}_{[l,j,\mathbf{U}_{[l-1,k,{\mathbf{C}}[l-1], {\mathbf{A}}[l-1]]},{\mathbf{C}}[l-1],
\mathbf{C}_{[l,i,{\mathbf{C}}[l-1]]}]})
\end{split}
\een
are jointly $\e_l$-typical with respect to $P_{\tilde{U}\tilde{A}\tilde{Y}\tilde{C}CAY}$ and $0$ otherwise. We have
\be
\begin{split}
 \text{Pr}(\mc{E}_4 | \mc{F})  & =  P\big( \exists \ (i,j,k) \neq (G_C[l],G_A[l],G_U[l-1]) \\
&  \qquad \text{ s.t. } \psi(i,j,k)=1 \mid \mc{F} \big) \\
& = \Phi_1+\Phi_2+\Phi_3+\Phi_4,
\end{split}
\label{eq:PE4_F}
\ee
where
\begin{gather}
\Phi_1  = \text{Pr}( \exists \ j \neq G_A[l] \text{ s.t. } \psi(G_C[l],j,G_U[l-1])=1 \mid \mc{F} ), \\
\Phi_2  = \text{Pr}( \exists \ i \neq G_C[l], j  \text{ s.t. } \psi(i,j,G_U[l-1])=1 \mid \mc{F} ), \\
\Phi_3  = \text{Pr}( \exists \ k \neq G_U[l-1], j \text{ s.t. } \psi(G_C[l],j,k)=1 \mid \mc{F} ), \\
\Phi_4  = \text{Pr}( \exists \ i \neq G_C[l], k \neq G_U[l-1], j \text{ s.t. } \psi(i,j,k)=1 | \mc{F}).
\end{gather}
\subsection{Upper bound for $\Phi_1$}
Using the union bound, we have
\be
\begin{split}
 \Phi_1 \leq  & \sum_{j=1}^{2^{n\rho_1}} \text{Pr}   \Big( \ \{\mathbf{U}[l-1], \mathbf{A}[l-1],\mathbf{Y}[l-1], \mathbf{C}[l-1], \mathbf{C}[l], \\
&  \ \mathbf{Y}[l], \mathbf{A}_{[l,j,\mathbf{U}[l-1],{\mathbf{C}}[l-1], \mathbf{C}[l]]} \}
\in \mathcal{A}^{(n)}_{\e_l}(P_{\tilde{C} \tilde{A} \tilde{U} \tilde{Y} CAY }), \\
& \  G_A[l] \neq j  \mid \mc{F} \Big).
\end{split}
\label{eq:phi1_1}
\ee
For brevity, we denote the tuple $(\mathbf{U}[l-1],{\mathbf{C}}[l-1], \mathbf{C}[l])$ by $\mathbf{T}'$ and the tuple $(\mathbf{U}[l-1],{\mathbf{C}}[l-1], \mathbf{C}[l],\mathbf{A}[l-1],\mathbf{Y}[l-1])$ by $\mathbf{T}$. \eqref{eq:phi1_1} can then be written as
\be
\begin{split}
& \Phi_1 \leq \frac{1}{\text{Pr}[\mc{F}]} \sum_{j=1}^{2^{n\rho_1}} \sum_{\mathbf{t}, \mathbf{a}, \mathbf{y} \in \mc{A}_{\e_l}}
\text{Pr}[ \mathbf{T} = \mathbf{t}, \ \mathbf{A}_{[l,j,\mathbf{T}']} = \mathbf{a}, \\
& \qquad \qquad \qquad \qquad \qquad \quad \mathbf{Y}[l]=\mathbf{y}, \ G_A[l] \neq j ]\\
&=\frac{2^{n \rho_1}}{\text{Pr}[\mc{F}]} \sum_{\mathbf{t}, \mathbf{a}, \mathbf{y} \in \mc{A}_{\e_l}}  \text{Pr}[\mathbf{T} = \mathbf{t}] \ \text{Pr}[ \mathbf{A}_{[l,1,\mathbf{T}']} = \mathbf{a}, \\
& \qquad \qquad \qquad \qquad \quad \mathbf{Y}[l]=\mathbf{y}, \ G_A[l] \neq 1  \mid  \mathbf{T}=\mathbf{t} ]
\end{split}
\label{eq:phi1_2}
\ee
where the second equality is due to the symmetry of the codebook construction. We note that the index $G_A[l]$ is a \emph{function} of the entire $A$-codebook
$\{  \mathbf{A}_{[l,j,\mathbf{U}[l-1],{\mathbf{C}}[l-1], \mathbf{C}[l]]}, \ j=1\ldots 2^{n\rho_1} \}$ and so conditioned on  $\mathbf{T}=\mathbf{t}$, the events
\[ G_A[l] \neq 1 \ \text{ and } \ ( \mathbf{A}_{[l,1,\mathbf{U}[l-1],{\mathbf{C}}[l-1], \mathbf{C}[l]]} = \mathbf{a}, \ \mathbf{Y}[l]=\mathbf{y})  \]
are dependent. This dependency can be handled using the technique developed in \cite{LimMinKim}.

Let $\mc{\bar{C}}$ be the set $\{ \mathbf{A}_{[l,j,\mathbf{U}[l-1],{\mathbf{C}}[l-1], \mathbf{C}[l]]}, \ j=2\ldots 2^{n\rho_1} \}$, i.e., $\mc{\bar{C}}$ is the $A$-codebook without the first codeword.
Focusing on the inner term of the summation in \eqref{eq:phi1_2}, we have
\be
\begin{split}
& \text{Pr}[\mathbf{A}_{[l,1,\mathbf{T}']} = \mathbf{a}, \ \mathbf{Y}[l]=\mathbf{y}, \ G_A[l] \neq 1  \mid  \mathbf{T}=\mathbf{t}] \\
& \leq \text{Pr}[\mathbf{A}_{[l,1,\mathbf{T}']} = \mathbf{a}, \ \mathbf{Y}[l]=\mathbf{y}  \mid G_A[l] \neq 1,  \mathbf{T}=\mathbf{t}] \\
& = \sum_{\bar{\emph{c}}} \text{Pr}[\mathbf{A}_{[l,1,\mathbf{T}']} = \mathbf{a}, \ \mathbf{Y}[l]=\mathbf{y}  \mid G_A[l] \neq 1,  \mathbf{T}=\mathbf{t}, \mc{\bar{C}} =\bar{\emph{c}}] \\
& \qquad \quad \cdot \text{Pr}[\mc{\bar{C}} =\bar{\emph{c}} \mid G_A[l] \neq 1,  \mathbf{T}=\mathbf{t}] \\
& \stackrel{(a)}{=} \sum_{\bar{\emph{c}}} \text{Pr}[\mathbf{A}_{[l,1,\mathbf{T}']} = \mathbf{a} \mid G_A[l] \neq 1,  \mathbf{T}=\mathbf{t}, \mc{\bar{C}} =\bar{\emph{c}}] \\
& \qquad \quad \cdot \text{Pr}[\mathbf{Y}[l]=\mathbf{y}  \mid G_A[l] \neq 1,  \mathbf{T}=\mathbf{t}, \mc{\bar{C}} =\bar{\emph{c}}] \\
& \qquad \quad \cdot \text{Pr}[\mc{\bar{C}} =\bar{\emph{c}} \mid G_A[l] \neq 1,  \mathbf{T}=\mathbf{t}] \\
&  \stackrel{(b)}{\leq}  2 \cdot \text{Pr}[\mathbf{A}_{[l,1,\mathbf{T}']} = \mathbf{a} \mid
\mathbf{T}' = \mathbf{t}'] \cdot \\
& \quad \ \sum_{\bar{\emph{c}}} \text{Pr}[\mathbf{Y}[l]=\mathbf{y}  \mid G_A[l] \neq 1,  \mathbf{T}=\mathbf{t}, \mc{\bar{C}} =\bar{\emph{c}}] \\
& \qquad \quad \cdot  \text{Pr}[\mc{\bar{C}} =\bar{\emph{c}} \mid G_A[l] \neq 1,  \mathbf{T}=\mathbf{t}]\\
& \stackrel{(c)}{\leq} 4 \cdot \text{Pr}[\mathbf{A}_{[l,1,\mathbf{T}']} = \mathbf{a} \mid
\mathbf{T}' = \mathbf{t}'] \ \text{Pr}[\mathbf{Y}[l]=\mathbf{y}  \mid \mathbf{T}=\mathbf{t}] \\
& \stackrel{(d)}{\leq}  4  \cdot  2^{-n(H(A| C \tilde{C} \tilde{U}) - \delta(\e_l))} \cdot
2^{-n(H(Y| C \tilde{C} \tilde{U} \tilde{A} \tilde{Y}) - \delta(\e_l) )}.
\end{split}
\label{eq:chain_phi1}
\ee
In the chain above, $(a)$ is true because given $G_A[l] \neq 1$, we have the Markov chain
$ \mathbf{A}_{[l,1,\mathbf{T}']} -(\mc{\bar{C}}, \mathbf{T}) - \mathbf{X}[l] - \mathbf{Y}[l].$
$(d)$ follows from Property $3$ of typical sequences, while $(b)$ and $(c)$ are obtained from the following claim, proved along the lines of \cite[Lemmas 1 and 2]{LimMinKim}.

\begin{claim}
\ben
\begin{split}
& \text{Pr}[\mathbf{A}_{[l,1,\mathbf{T}']} = \mathbf{a} \mid G_A[l] \neq 1,  \mathbf{T}=\mathbf{t}, \mc{\bar{C}} =\bar{\emph{c}}] \\
& \quad \leq  \ 2 \; \text{Pr}[\mathbf{A}_{[l,1,\mathbf{T}']} = \mathbf{a} \mid  \mathbf{T}'=\mathbf{t}'], \ \text{ and} \\
& \text{Pr}[\mathbf{Y}[l]=\mathbf{y}  \mid G_A[l] \neq 1,  \mathbf{T}=\mathbf{t}] \ \leq \ 2 \;\text{Pr}[\mathbf{Y}[l]=\mathbf{y}  \mid  \mathbf{T}=\mathbf{t}].
\end{split}
\een
\label{claim1}
\end{claim}
\begin{IEEEproof}
We have
\be
\begin{split}
&\text{Pr}[\mathbf{A}_{[l,1,\mathbf{T}']} = \mathbf{a} \mid G_A[l] \neq 1,  \mathbf{T}=\mathbf{t}, \mc{\bar{C}} =\bar{\emph{c}}]\\
&= \frac{ \text{Pr}[\mathbf{A}_{[l,1,\mathbf{T}']} = \mathbf{a},   G_A[l] \neq 1  \mid \mathbf{T}=\mathbf{t}, \mc{\bar{C}} =\bar{\emph{c}}] }
{\text{Pr}[ G_A[l] \neq 1,  \mid \mathbf{T}=\mathbf{t}, \mc{\bar{C}} =\bar{\emph{c}}]} \\
&\leq \frac{ \text{Pr}[\mathbf{A}_{[l,1,\mathbf{T}']} = \mathbf{a} \mid \mathbf{T}=\mathbf{t}, \mc{\bar{C}} =\bar{\emph{c}}] }{\text{Pr}[ G_A[l] \neq 1 \mid \mathbf{T}=\mathbf{t}, \mc{\bar{C}} =\bar{\emph{c}}]}\\
&= \frac{ \text{Pr}[\mathbf{A}_{[l,1,\mathbf{T}']} = \mathbf{a} \mid \mathbf{T}'=\mathbf{t}'] }{\text{Pr}[ G_A[l] \neq 1 \mid \mathbf{T}=\mathbf{t}, \mc{\bar{C}} =\bar{\emph{c}}]}
\end{split}
\label{eq:claimA}
\ee
where the last equality holds because each codeword of the codebook $\{ \mathbf{A}_{[l,j,\mathbf{T}']} , \ j=1,\ldots 2^{n\rho_1} \}$ is independently generated, conditioned only on the symbols of $\mathbf{T}'$. We now provide a lower bound for the denominator of \eqref{eq:claimA}.
\be
\begin{split}
&\text{Pr}[ G_A[l] \neq 1 \mid \mathbf{T}=\mathbf{t}, \mc{\bar{C}} =\bar{\emph{c}}] \\
&= 1 -  \text{Pr}[ G_A[l] = 1 \mid \mathbf{T}=\mathbf{t}, \mc{\bar{C}} =\bar{\emph{c}}]\\
&\geq 1 - \text{Pr}\left[(\mathbf{A}_{[l,1,\mathbf{T}']}, \mathbf{T}) \in \mathcal{A}^{(n)}_{\e_l}(P_{\tilde{C} \tilde{A} \tilde{U} \tilde{Y} CA }) \right] \\
& \geq 1- 2^{-n(I(A; \tilde{A} \tilde{Y}| \tilde{U} \tilde{C}  C ) - \delta(\e_l))}  \geq \frac{1}{2}
\end{split}
\ee
for sufficiently large $n$. Substituting in \eqref{eq:claimA} completes the proof of the first part of the claim.

For the second part, we write
\be
\begin{split}
&\text{Pr}[\mathbf{Y}[l]=\mathbf{y} \mid G_A[l] \neq 1,  \mathbf{T}=\mathbf{t}] \\
%
&\leq \frac{ \text{Pr}[\mathbf{Y}[l]=\mathbf{y} \mid \mathbf{T}=\mathbf{t}] }{\text{Pr}[ G_A[l] \neq 1 \mid \mathbf{T}=\mathbf{t}]}\\
&= \frac{ \text{Pr}[\mathbf{Y}[l]=\mathbf{y} \mid \mathbf{T}=\mathbf{t}] }{(2^{n\rho_1} - 1)/2^{n\rho_1}}
 \leq \ 2 \cdot \text{Pr}[\mathbf{Y}[l]=\mathbf{y} \mid \mathbf{T}=\mathbf{t}]
\end{split}
\label{eq:claimB}
\ee
for large enough $n$. The equality above is due to the symmetry of the codebook construction. The claim is proved.
\end{IEEEproof}

Substituting the bound from \eqref{eq:chain_phi1} in \eqref{eq:phi1_2}, we obtain
\be
\begin{split}
 \Phi_1  &\leq  \frac{2^{n \rho_1}}{\text{Pr}[\mc{F}]} \sum_{\mathbf{t}}  \text{Pr}[\mathbf{T} = \mathbf{t} ]  \sum_{\mathbf{a}, \mathbf{y} \in \mc{A}_{\e_l}(. |\mathbf{t})} \hspace{-5pt} 4 \cdot 2^{2n \delta(\e_l)}   2^{-nH(A| C \tilde{C} \tilde{U})} \\
& \qquad \qquad \qquad \qquad \qquad \qquad  \cdot 2^{-nH(Y| C \tilde{C} \tilde{U} \tilde{A} \tilde{Y})} \\
& \stackrel{(a)}{\leq} \frac{2^{n \rho_1}}{\text{Pr}[\mc{F}]} \sum_{\mathbf{t}}  \text{Pr}[\mathbf{T} = \mathbf{t}] \ 2^{n (H(AY| C \tilde{C} \tilde{U} \tilde{A} \tilde{Y}) + \delta(\e_l))} \\
 & \qquad \quad \cdot \left(4 \cdot 2^{2n \delta(\e_l)} \cdot 2^{-nH(A| C \tilde{C} \tilde{U})} \cdot 2^{-nH(Y| C \tilde{C} \tilde{U} \tilde{A} \tilde{Y})} \right) \\
& {=} \  \frac{4 \cdot 2^{3n \delta(\e_l)} \cdot 2^{n \rho_1} \cdot 2^{n H(AY| C \tilde{C} \tilde{U} \tilde{A} \tilde{Y})} }
{ \text{Pr}[\mc{F}] \cdot 2^{nH(A| C \tilde{C} \tilde{U})} \cdot 2^{nH(Y| C \tilde{C} \tilde{U} \tilde{A} \tilde{Y})}}
\end{split}
\label{eq:Phi1_final}
\ee
where $(a)$ follows from the upper bound on the size of the conditionally typical set (Property $2$).
\subsection{Upper bound for $\Phi_2$, $\Phi_3$, $\Phi_4$}
Using the union bound, we have
\be
\begin{split}
&\Phi_2 \leq \sum_{i=1}^{2^{n\rho_0}} \sum_{j=1}^{2^{n\rho_1}} \text{Pr} \Big( \{\mathbf{U}[l-1], \mathbf{A}[l-1],\mathbf{Y}[l-1], \mathbf{C}[l-1],\\
& \qquad \qquad  \mathbf{Y}[l], \mathbf{C}_{[l, i, \mathbf{C}[l-1]]}, \mathbf{A}_{[l,j,\mathbf{U}[l-1],{\mathbf{C}}[l-1], \mathbf{C}_{[l, i, \mathbf{C}[l-1]]}]}\} \\
 & \qquad \qquad  \in \mathcal{A}^{(n)}_{\e_l}(P_{\tilde{C} \tilde{A} \tilde{U} \tilde{Y} CAY }), \ G_C[l] \neq i  \mid \mc{F}\Big).
\label{eq:phi2_1}
\end{split}
\ee

To keep the notation manageable, in the next few equations we will use the shorthand $\mathbf{C}_{i}$ for $\mathbf{C}_{[l, i, \mathbf{C}[l-1]]}$.
We also redefine $\mathbf{T}'$ as the tuple $(\mathbf{U}[l-1],{\mathbf{C}}[l-1])$ and $\mathbf{T}$ as the tuple $(\mathbf{U}[l-1],{\mathbf{C}}[l-1], \mathbf{A}[l-1],\mathbf{Y}[l-1])$.  \eqref{eq:phi2_1} can then be written as
\be
\begin{split}
& \Phi_2 \leq  \frac{1}{\text{Pr}[\mc{F}]} \sum_{i=1}^{2^{n\rho_0}} \sum_{j=1}^{2^{n\rho_1}} \sum_{\mathbf{t}, \mathbf{c}, \mathbf{a}, \mathbf{y} \in \mc{A}_{\e_l}}
\text{Pr}[ \mathbf{T} = \mathbf{t}, \mathbf{C}_i=\mathbf{c},    \\
 & \qquad \qquad \qquad \qquad \mathbf{A}_{[l,j,\mathbf{T}', \mathbf{C}_i]} = \mathbf{a}, \mathbf{Y}[l]=\mathbf{y}, \ G_C[l] \neq i ]\\
&=\frac{2^{n (\rho_0+\rho_1)}}{\text{Pr}[\mc{F}]} \sum_{\mathbf{t}, \mathbf{a},  \mathbf{c}, \mathbf{y} \in \mc{A}_{\e_l}}  \text{Pr}[\mathbf{T} = \mathbf{t}]
\cdot \text{Pr}[ \mathbf{C}_1=\mathbf{c}, \\
 & \qquad \qquad \mathbf{A}_{[l,1,\mathbf{T}', \mathbf{C}_1]} = \mathbf{a}, \ \mathbf{Y}[l]=\mathbf{y}, \ G_C[l] \neq 1  \mid  \mathbf{T}=\mathbf{t} ]
\end{split}
\label{eq:phi2_2}
\ee
where the second equality is due to the symmetry of the codebook construction. We note that the index $G_C[l]$ is a {function} of the entire $C$-codebook
$\{  \mathbf{C}_i = \mathbf{C}_{[l, i, \mathbf{C}[l-1]]}, \ i=1\ldots 2^{n\rho_0} \}$ and so conditioned on  $\mathbf{T}=\mathbf{t}$, the events
 $G_C[1] \neq 1$ and
\[ ( \mathbf{C}_1= \mathbf{c}, \mathbf{A}_{[l,1,\mathbf{U}[l-1], \mathbf{C}[l-1], \mathbf{C}_1]} = \mathbf{a}, \ \mathbf{Y}[l]=\mathbf{y})  \]
are dependent.
Define $\mc{\bar{C}}$  as $\{\mathbf{C}_i = \mathbf{C}_{[l, i, \mathbf{C}[l-1]]}, \ i=2\ldots 2^{n\rho_0} \}$, i.e., the $C$-codebook without the first codeword. We then have
\be
\begin{split}
& \text{Pr}[\mathbf{C}_1=\mathbf{c}, \mathbf{A}_{[l,1,\mathbf{T}', \mathbf{C}_1]} = \mathbf{a},  \mathbf{Y}[l]=\mathbf{y}, \ G_C[l] \neq 1  \mid  \mathbf{T}=\mathbf{t}] \\
& \leq \text{Pr}[\mathbf{C}_1=\mathbf{c}, \mathbf{A}_{[l,1,\mathbf{T}', \mathbf{C}_1]} = \mathbf{a}, \mathbf{Y}[l]=\mathbf{y} | G_C[l] \neq 1,  \mathbf{T}=\mathbf{t}] \\
& = \sum_{\bar{\emph{c}}} \text{Pr}[\mathbf{C}_1=\mathbf{c}, \mathbf{A}_{[l,1,\mathbf{T}']} = \mathbf{a},  \mathbf{Y}[l]=\mathbf{y}  \mid G_C[l] \neq 1,  \\
& \qquad \quad  \mathbf{T}=\mathbf{t}, \mc{\bar{C}} =\bar{\emph{c}}]  \cdot \text{Pr}[\mc{\bar{C}} =\bar{\emph{c}} \mid G_C[l] \neq 1,  \mathbf{T}=\mathbf{t}] \\
& \stackrel{(a)}{=} \sum_{\bar{\emph{c}}} \text{Pr}[\mathbf{C}_1=\mathbf{c}, \mathbf{A}_{[l,1,\mathbf{T}']} = \mathbf{a} \mid G_C[l] \neq 1,  \mathbf{T}=\mathbf{t}, \mc{\bar{C}} =\bar{\emph{c}}] \\
& \qquad \quad  \cdot \text{Pr}[\mathbf{Y}[l]=\mathbf{y}  \mid G_C[l] \neq 1,  \mathbf{T}=\mathbf{t}, \mc{\bar{C}} =\bar{\emph{c}}] \\
& \qquad \quad \cdot \text{Pr}[\mc{\bar{C}} =\bar{\emph{c}} \mid G_C[l] \neq 1,  \mathbf{T}=\mathbf{t}] \\
&  \stackrel{(b)}{\leq}  2  \cdot \text{Pr}[\mathbf{C}_1=\mathbf{c}, \mathbf{A}_{[l,1,\mathbf{T}']} = \mathbf{a}  \mid
\mathbf{T}' = \mathbf{t}']
\sum_{\bar{\emph{c}}} \text{Pr}[\mathbf{Y}[l]=\mathbf{y}  \\
&  \quad \mid  G_A[l] \neq 1, \mathbf{T}=\mathbf{t}, \mc{\bar{C}} =\bar{\emph{c}}] \ \text{Pr}[\mc{\bar{C}} =\bar{\emph{c}} \mid G_A[l] \neq 1,  \mathbf{T}=\mathbf{t}]\\
& \stackrel{(c)}{\leq} 4  \cdot \text{Pr}[\mathbf{C}_1=\mathbf{c}, \mathbf{A}_{[l,1,\mathbf{T}']} = \mathbf{a} |
\mathbf{T}' = \mathbf{t}'] \ \text{Pr}[\mathbf{Y}[l]=\mathbf{y}  | \mathbf{T}=\mathbf{t}] \\
& \stackrel{(d)}{\leq}  4 \cdot  2^{-n(  H(C|\tilde{C}) + H(A| C \tilde{C} \tilde{U}) - \delta(\e_l) )}
\cdot 2^{-n(H(Y| \tilde{C} \tilde{U} \tilde{A} \tilde{Y}) - \delta(\e_l) )}.
\end{split}
\label{eq:chain_phi2}
\ee
 Given $G_C[l] \neq 1$, $(a)$ is true because we have the Markov chain
$ (\mathbf{C}_{[l, 1, \mathbf{C}[l-1]]}, \mathbf{A}_{[l,1,\mathbf{T}']}) -(\mc{\bar{C}}, \mathbf{T}) - \mathbf{X}[l] - \mathbf{Y}[l]$.
$(d)$ follows from Property $3$ of typical sequences, while $(b)$ and $(c)$ follow from arguments very similar to
Claim \ref{claim1}.
Substituting the bound from \eqref{eq:chain_phi2} in \eqref{eq:phi2_2}, we obtain
\be
\begin{split}
& \Phi_2  \leq  \frac{2^{n(\rho_0 + \rho_1)}}{\text{Pr}[\mc{F}]} \sum_{\mathbf{t}}  \text{Pr}[\mathbf{T} = \mathbf{t} ]
 \sum_{\mathbf{c}, \mathbf{a}, \mathbf{y} \in \mc{A}_{\e_l}(. |\mathbf{t})} \Big(4 \cdot 2^{2n \delta(\e_l)} \\
& \qquad \cdot  2^{-nH(C|\tilde{C})}  \cdot  2^{-nH(A| C \tilde{C} \tilde{U})} \cdot 2^{-nH(Y| \tilde{C} \tilde{U} \tilde{A} \tilde{Y})} \Big).
\end{split}
\ee
 Using the upper bound for the size of the conditionally typical set, we have
\be
\begin{split}
\Phi_2 & \leq \frac{2^{n(\rho_0 + \rho_1)}}{\text{Pr}[\mc{F}]} \sum_{\mathbf{t}}  \text{Pr}[\mathbf{T} = \mathbf{t}] \
2^{n (H(AYC | \tilde{C} \tilde{U} \tilde{A} \tilde{Y}) + \delta(\e_l))}\\
& \ \left(4 \cdot 2^{2n \delta(\e_l)} \cdot  2^{-nH(C|\tilde{C})} \cdot  2^{-nH(A| C \tilde{C} \tilde{U})} \cdot 2^{-nH(Y|  \tilde{C} \tilde{U} \tilde{A} \tilde{Y})} \right) \\
& {=} \  \frac{4 \cdot 2^{3n \delta(\e_l)} \cdot 2^{n (\rho_1 + \rho_0)} \cdot 2^{n H(AYC | \tilde{C} \tilde{U} \tilde{A} \tilde{Y})} }
{ \text{Pr}[\mc{F}] \cdot  2^{nH(C|\tilde{C})} \cdot  2^{nH(A| C \tilde{C} \tilde{U})} \cdot 2^{nH(Y|  \tilde{C} \tilde{U} \tilde{A} \tilde{Y})} }
\end{split}
\label{eq:Phi2_final}
\ee

In a similar fashion, we can obtain the following bounds for $\Phi_3$ and $\Phi_4$.
\be
\Phi_3  \leq \frac{4 \cdot 2^{3n \delta(\e_l)} \cdot 2^{n (R'_1+\rho_1)} \cdot 2^{n H(\tilde{U} AY | C \tilde{C}  \tilde{A} \tilde{Y})} }
{ \text{Pr}[\mc{F}] \cdot  2^{nH(\tilde{U}| \tilde{A} \tilde{C})} \cdot  2^{nH(A| C \tilde{C} \tilde{U})} \cdot 2^{nH(Y| C \tilde{C} \tilde{A} \tilde{Y})} },
\label{eq:Phi3_final}
\ee
\be
\Phi_4  \leq \frac{4 \cdot 2^{3n \delta(\e_l)} \cdot 2^{n (R'_1+\rho_0+\rho_1)} \cdot 2^{n H(\tilde{U}C AY |  \tilde{C}  \tilde{A} \tilde{Y})} }
{ \text{Pr}[\mc{F}] \ 2^{nH(\tilde{U}| \tilde{A} \tilde{C})} 2^{nH(C|\tilde{C})}   2^{nH(A| C \tilde{C} \tilde{U})}  2^{nH(Y|\tilde{C} \tilde{A} \tilde{Y})} }.
\label{eq:Phi4_final}
\ee

Lemmas \ref{lem:error1} and \ref{lem:error2}  together with the induction hypothesis that $\text{Pr}[\mc{E}_4[k]] < \epsilon$ for $k=1, \ldots, l-1$ imply that
$\text{Pr}[\mc{F}] > 1- 5\epsilon l$, which is close to $1$ for $\epsilon \ll 1/L$. Thus the bounds \eqref{eq:Phi1_final}, \eqref{eq:Phi2_final}, \eqref{eq:Phi3_final} and \eqref{eq:Phi4_final} can be made arbitrarily small for  sufficiently large $n$ if the conditions of the lemma are satisfied.

Substituting back in \eqref{eq:PE4_F}, we obtain $P[\mc{E}_4[l] \mid \mc{F}] \leq {\e}$ for all sufficiently large $n$.
Similarly, one can show that $P[\mc{E}_5[l] \mid \mc{F}]  \leq {\e}$  if the conditions in the  lemma are satisfied.

\IEEEtriggeratref{8}

\section*{Acknowledgement}
We thank the associate editor and the anonymous reviewers for their comments and suggestions, which led to a much improved paper.


\begin{thebibliography}{31}

\bibitem{Marton79}
K.~Marton, ``A coding theorem for the discrete memoryless broadcast channel,''
  {\em IEEE Trans. Inf. Theory}, vol.~25, no.~3, pp.~306--311, 1979.

\bibitem{Cover98}
T.~M. Cover, ``Comments on broadcast channels,'' {\em IEEE Trans. Inf. Theory},
  vol.~44, no.~6, pp.~2524--2530, 1998.

\bibitem{ElG78}
A.~{E}l {G}amal, ``The feedback capacity of degraded broadcast channels,'' {\em
  IEEE Trans. Inf. Theory}, vol.~24, pp.~379--381, May 1980.

\bibitem{Dueck80}
G.~Dueck, ``Partial feedback for two-way and broadcast channels,'' {\em Inform.
  and Control}, vol.~46, pp.~1--15, July 1980.

\bibitem{OzarowLeung84}
L.~H. Ozarow and S.~K. Leung-Yan-Cheong, ``An achievable region and outer bound
  for the {G}aussian broadcast channel with feedback,'' {\em IEEE Trans.
  Inf. Theory}, vol.~30, no.~4, pp.~667--671, 1984.

\bibitem{Bhaskaran08}
S.~R. Bhaskaran, ``Gaussian broadcast channel with feedback,'' {\em IEEE Trans.
    Inf. Theory}, vol.~54, no.~11, pp.~5252--5257, 2008.

\bibitem{Kramer03}
G.~Kramer, ``{Capacity Results for the Discrete Memoryless Network},'' {\em
  IEEE Trans. Inf. Theory}, vol.~49, pp.~4--20, January 2003.

\bibitem{HanCosta87}
T.~S. Han and M.~H.~M. Costa, ``Broadcast channels with arbitrarily correlated
  sources,'' {\em IEEE Trans. Inf. Theory}, vol.~IT-33, no.~5, pp.~641--650,
  1987.

\bibitem{KramerNair09}
G.~Kramer and C.~Nair, ``Comments on {`B}roadcast channels with arbitrarily
  correlated sources{'},'' in {\em Proc. IEEE Int. Symp. on Inf. Theory}, June
  2009.

\bibitem{MineroKim09}
P.~Minero and Y.-H. Kim, ``Correlated sources over broadcast channels,'' in
  {\em Proc. IEEE Int. Symp. Inf. Theory}, June 2009.

\bibitem{ChoiP08}
S.~Choi and S.~S. Pradhan, ``A graph-based framework for transmission of
  correlated sources over broadcast channels,'' {\em IEEE Trans. Inf. Theory},
  vol.~54, no.~7, pp.~2841--2856, 2008.

\bibitem{KangKramer08}
W.~Kang and G.~Kramer, ``Broadcast channel with degraded source random
  variables and receiver side information,'' in {\em Proc. IEEE Int. Symp. on
  Inf. Theory}, June 2008.

\bibitem{PrelogDouble}
M.~Gastpar, A.~Lapidoth, Y.~Steinberg, and M.~A. Wigger, ``Feedback can double
  the prelog of some memoryless gaussian networks,''
\newblock http://arxiv.org/abs/1003.6082.

\bibitem{RP10}
R.~Venkataramanan and S.~S. Pradhan, ``Achievable rates for the broadcast
  channel with feedback,'' {\em Proc. IEEE Int. Symp. on Inf. Theory}, 2010.

\bibitem{WiggerShay}
O.~Shayevitz and M.~A. Wigger, ``On the capacity of the discrete memoryless
  broadcast channel with feedback,'' {\em Proc. IEEE Int. Symp. Inf. Theory},
  2010.

\bibitem{WiggerShayJourn}
O.~Shayevitz and M.~A. Wigger, ``On the capacity of the discrete memoryless
  broadcast channel with feedback,'' {\em IEEE Trans. Inf. Theory}, vol.~59,
  pp.~1329--1345, March 2013.

\bibitem{ElGMuelen81}
A.~{El Gamal} and E.~C. van~der Muelen, ``A proof of {M}arton's coding theorem
  for the discrete memoryless broadcast channel,'' {\em IEEE Trans. Inf.
  Theory}, vol.~IT-27, no.~1, pp.~120--122, 1981.

\bibitem{ElGKimBook}
A.~E. Gamal and Y.-H. Kim, {\em Network Information Theory}.
\newblock Cambridge University Press, 2012.

\bibitem{Han84}
T.~S. Han, ``A general coding scheme for the two-way channel,'' {\em IEEE
  Trans. Inf. Theory}, vol.~30, no.~1, pp.~35--43, 1984.

\bibitem{RamjiMAC11}
R.~Venkataramanan and S.~S. Pradhan, ``A new achievable rate region for the
  multiple-access channel with noiseless feedback,'' {\em IEEE Trans. Inf.
  Theory}, vol.~57, pp.~8038--8054, Dec. 2011.

\bibitem{SchalkwijkKailath66}
J.~P.~M. Schalkwijk and T.~Kailath, ``A coding scheme for additive noise
  channels with feedback- part {I}{I}: {B}and-limited signals,'' {\em IEEE
  Trans. on Information Theory}, vol.~IT-12, pp.~183--189, April 1966.

\bibitem{KimLapWeiss}
Y.-H. Kim, A.~Lapidoth, and T.~Weissman, ``The gaussian channel with noisy
  feedback,'' in {\em IEEE Int. Symp. Inf. Theory}, June 2007.

\bibitem{Cover72}
T.~Cover, ``Broadcast channels,'' {\em IEEE Trans. Inf. Theory}, vol.~18,
  pp.~2 -- 14, Jan 1972.

\bibitem{Bergmans74}
P.~Bergmans, ``A simple converse for broadcast channels with additive white
  gaussian noise (corresp.),'' {\em IEEE Trans. Inf. Theory}, vol.~20,
  pp.~279 -- 280, Mar 1974.

\bibitem{Costa84}
M.~Costa, ``Writing on dirty paper,'' {\em IEEE Trans. Inf. Theory}, vol.~29,
  pp.~439 -- 441, May 1983.

\bibitem{GelfPin}
S.~I. Gelfand and M.~S. Pinsker, ``Coding for channel with random parameters,''
  {\em Problems of Control and Information Theory}, vol.~9, no.~1, pp.~19 --
  31, 1980.

\bibitem{GrayWyner}
R.~M. Gray and A.~D. Wyner, ``Source coding for a simple network,'' {\em Bell
  System Technical Journal}, vol.~53, pp.~1681 -- 1721, Nov. 1974.

\bibitem{OrlitskyRoche}
A.~Orlitsky and J.~Roche, ``Coding for computing,'' {\em IEEE Trans. Inf.
  Theory}, vol.~47, pp.~903 -- 917, Mar 2001.

\bibitem{TianDS11}
C.~Tian, S.~N. Diggavi, and S.~Shamai, ``Approximate characterizations for the
  gaussian source broadcast distortion region,'' {\em IEEE Trans. Inf. Theory},
  vol.~57, no.~1, pp.~124--136, 2011.

\bibitem{BrossLT10}
S.~I. Bross, A.~Lapidoth, and S.~Tinguely, ``Broadcasting correlated
  gaussians,'' {\em IEEE Trans. Inf. Theory}, vol.~56, no.~7,
  pp.~3057--3068, 2010.

\bibitem{LimMinKim}
S.~H. Lim, P.~Minero, and Y.-H. Kim, ``Lossy communication of correlated
  sources over multiple access channels,'' in {\em $48$th Allerton Conf. on
  Communication, Control, and Computing}, pp.~851 --858, 2010.

\end{thebibliography}
\end{document}